\documentclass[timestamp]{acmart}

\AtBeginDocument{%
  \providecommand\BibTeX{{%
    \normalfont B\kern-0.5em{\scshape i\kern-0.25em b}\kern-0.8em\TeX}}}

\begin{document}

\title{Siting thousands of radio transmitter towers on terrains with billions of points}


\author{W. Randolph Franklin}
\email{mail@wrfranklin.org}
\orcid{1234-5678-9012xxxxxxxxx}
\affiliation{%
  \institution{Electrical, Computer, and Systems Engineering Dept., Rensselaer Polytechnic Institute}
  \streetaddress{110 8th St}
  \city{Troy}
  \state{NY}
  \country{USA}
  \postcode{12180}
}

\author{Salles Viana Gomes de Magalhães}
\email{sallesviana@gmail.com}
\affiliation{%
  \institution{Departamento de Informática, Universidade Federal de Viçosa}
  \streetaddress{Campus Universitário}
  \city{Viçosa}
  \state{MG}
  \postcode{36570-900}
  \country{Brasil}
}

\author{Wenli Li}
\email{Wenli.Li@microsoft.com}
\affiliation{%
  \institution{Microsoft Corp}
  \city{San Francisco}
  \state{CA}
  \country{USA}}

\renewcommand{\shortauthors}{Franklin, Magalhães, Li}

\begin{abstract}

This paper presents a system that sites (finds optimal locations for) thousands of radio transmitter towers on terrains of up to two billion elevation posts.  Applications include cellphone towers, camera systems, or even mitigating environmental visual nuisances.  The transmitters and receivers may be situated above the terrain.  The system has been parallelized with OpenMP to run on a multicore CPU.
  
\end{abstract}

\ccsdesc[500]{Applied computing~Cartography}
\ccsdesc[300]{Human-centered computing~Geographic visualization}
\ccsdesc[100]{Theory of computation~Computational geometry}
\ccsdesc[300]{Networks~Location based services}

\keywords{terrain visibility, viewshed, multiple observer siting, large terrain datasets}

\begin{teaserfigure}
\centering
  \fbox{\includegraphics[width=.18\textwidth]{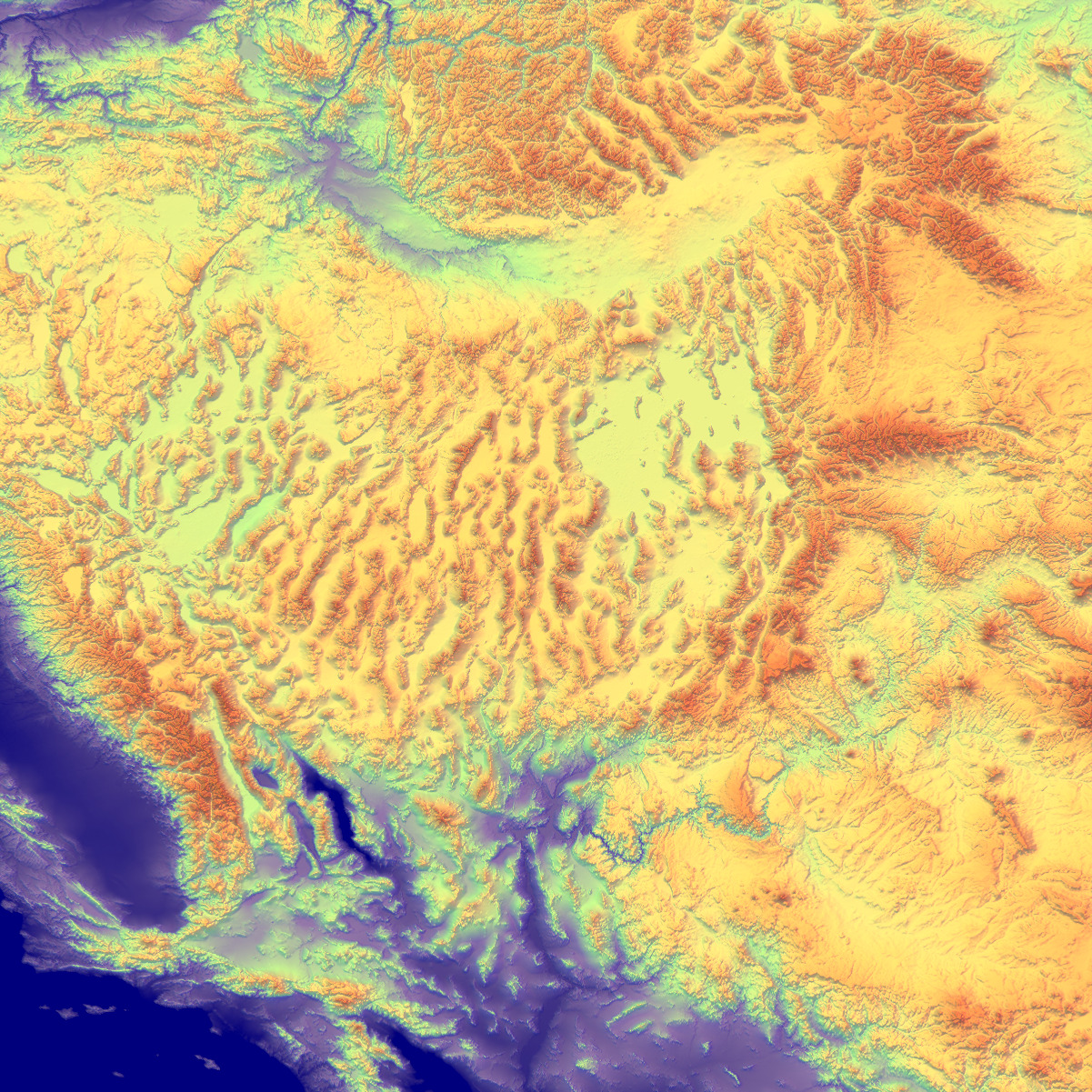}}
  \fbox{\includegraphics[width=.18\textwidth]{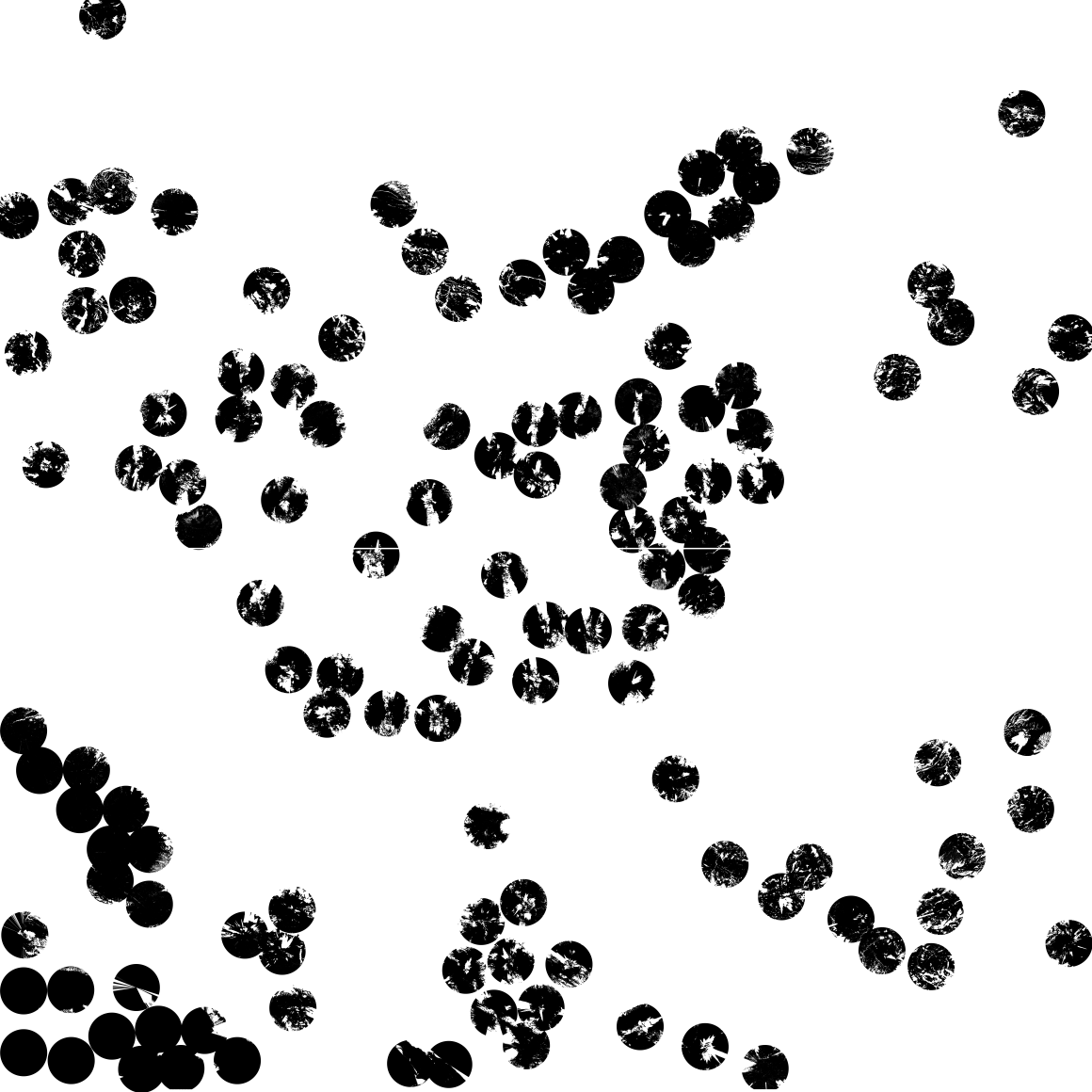}}
  \fbox{\includegraphics[width=.18\textwidth]{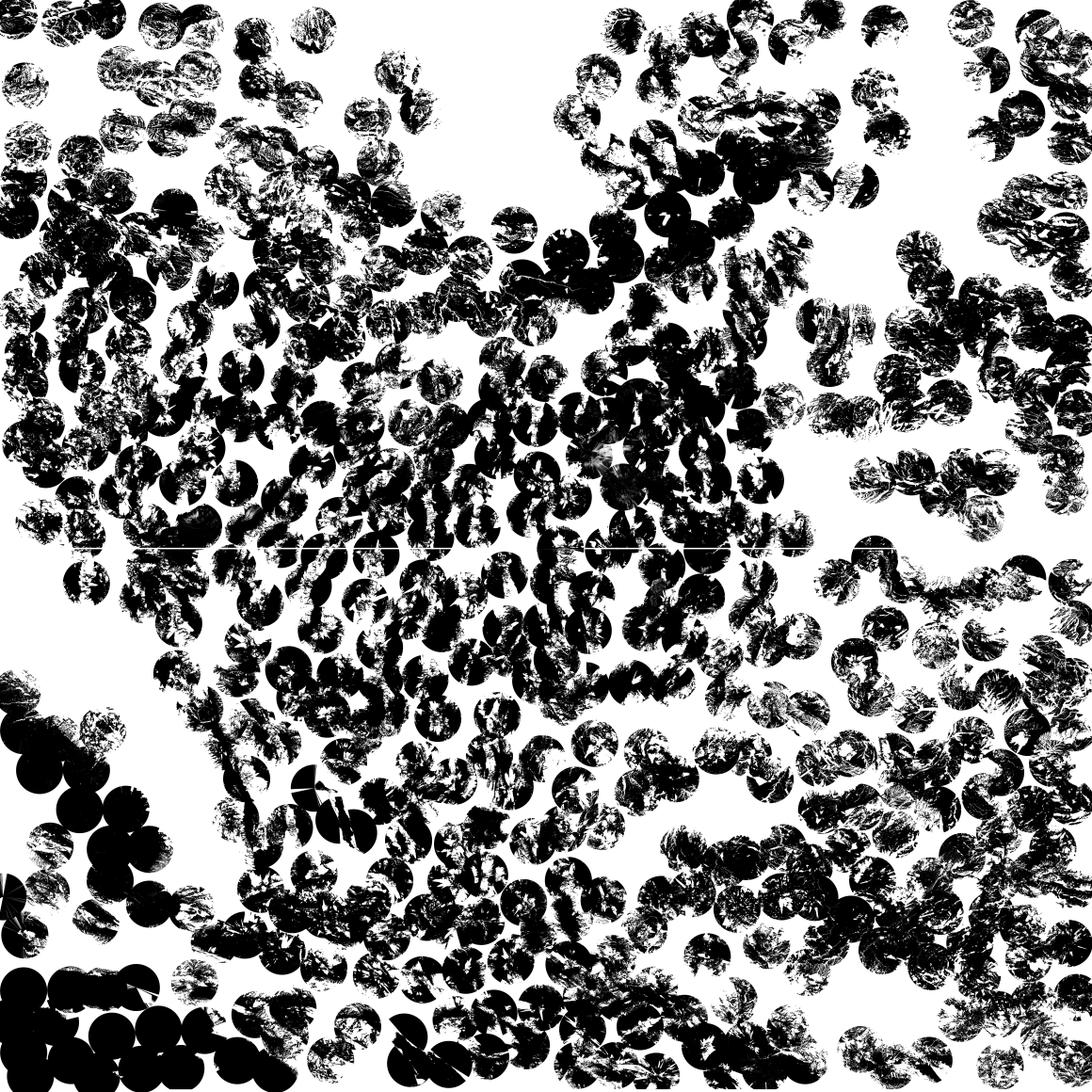}}
  \fbox{\includegraphics[width=.18\textwidth]{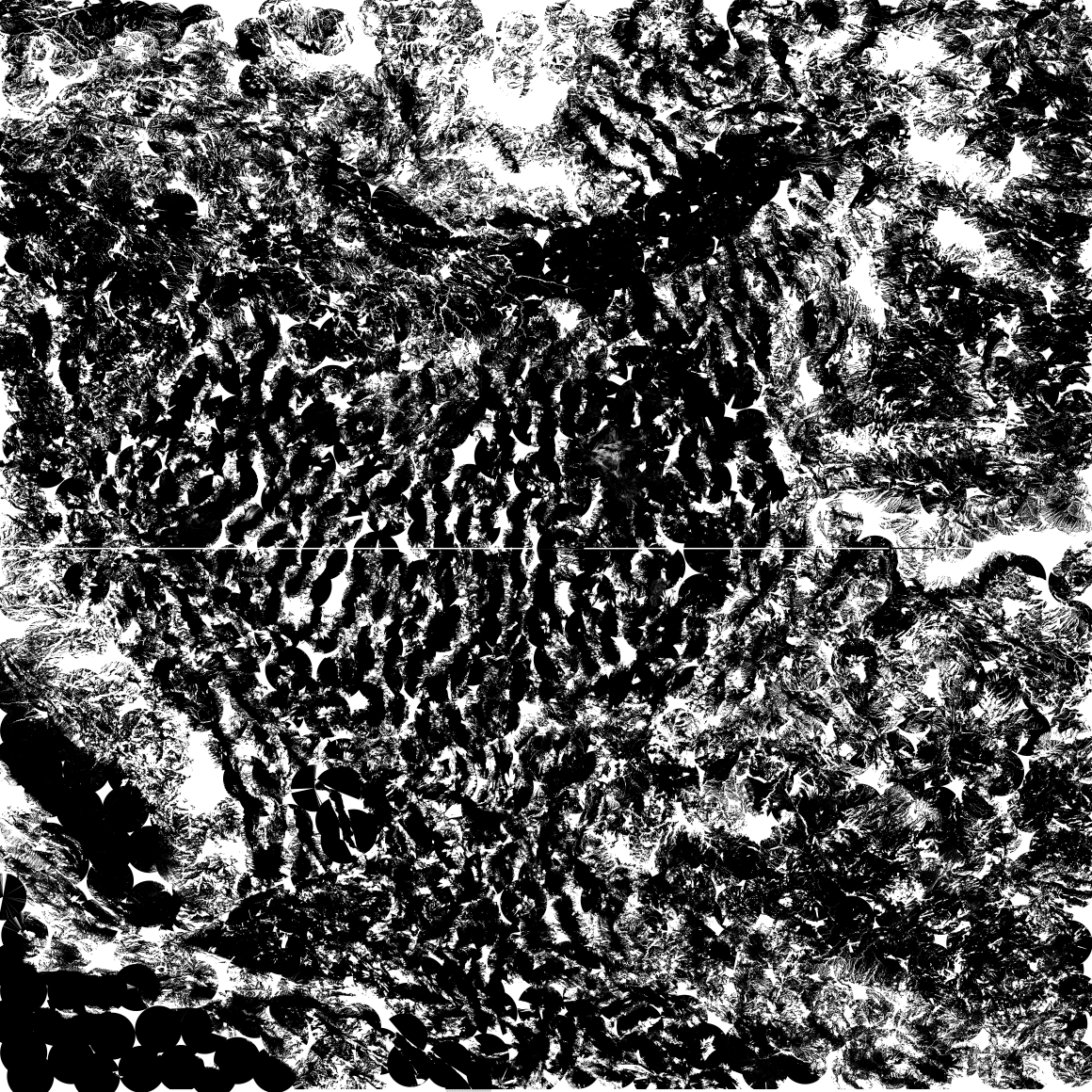}}
  \fbox{\includegraphics[width=.18\textwidth]{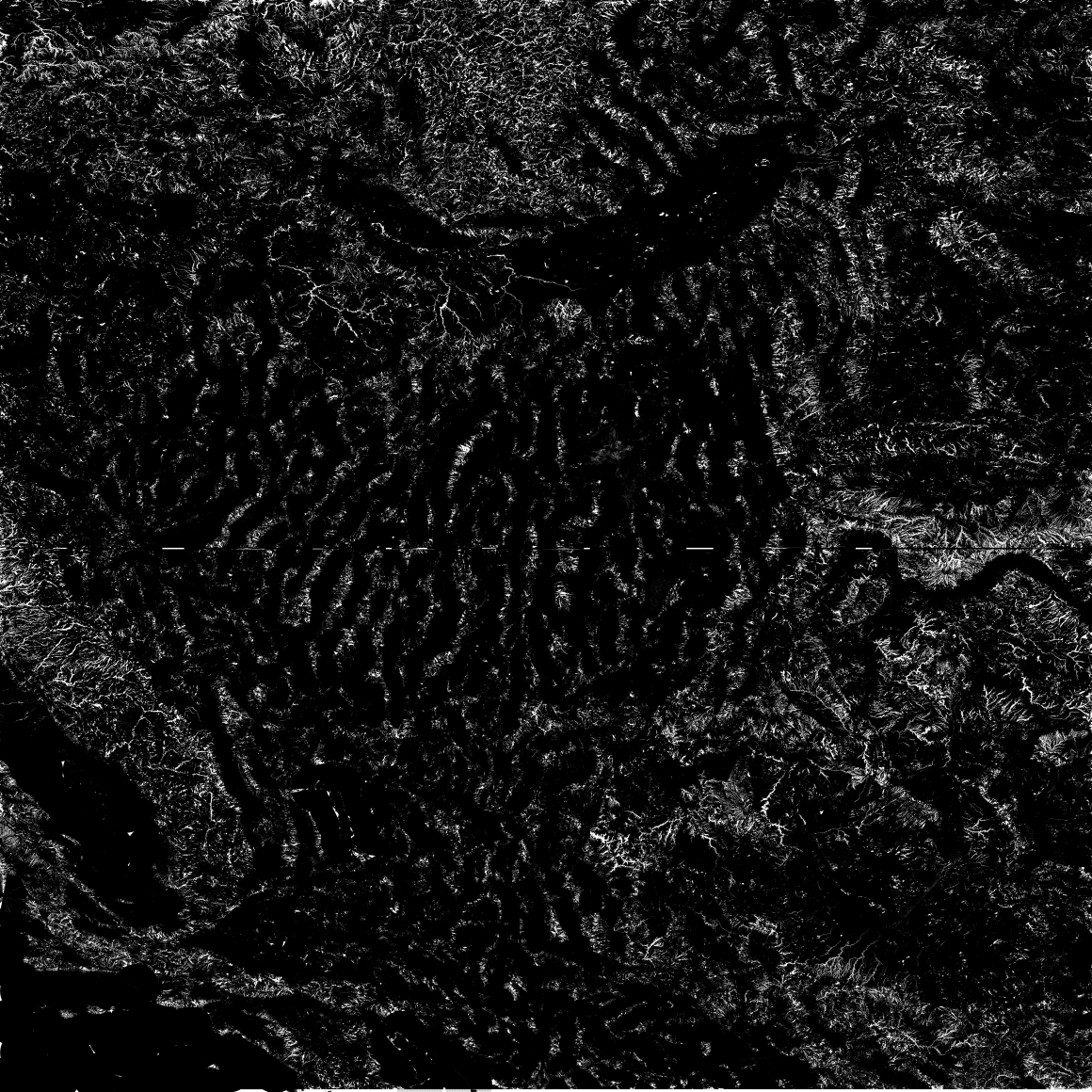}}
  \caption{US West data ($46400\times46400$ elevation posts); Cumulative viewsheds after siting 128, 512,  1024 and 4096 observers.} %
  \Description{Cumulative viewsheds after siting observers on the US West data.}
  \label{fig:teaser}
\end{teaserfigure}

\maketitle

\section{Definitions}

\begin{description}

\item[Terrain:] a single valued function $z(x,y)$ describing a land or water surface, with $(x,y)$ varying over some domain, typically a square.  The representation of this function will be discussed later.

\item[Transmitter:] a 3D point $(t_x, t_y, t_z)$ somewhere over the terrain; a source of straight-line radio or light waves.   There may be thousands of transmitters.

\item[Transmitter base:]  $(t_x, t_y, z(t_x, t_y))$ the point on the terrain directly below a transmitter.

\item[Transmitter height:] $h_t$, the vertical distance between a transmitter and its base.  Although this is not conceptually required, for simplicity, all the transmitters have the same height.

\item[Radius of interest:] ROI, the maximum distance that a transmitter can transmit to.   This is measured horizontally in 2D, not slantwise in 3D, and ignores possible differing elevations of the transmitter and receiver.

\item[Receiver:]  a 3D point $(r_x, r_y, r_z)$  somewhere over the terrain, which is intended to receive a signal from a transmitter.   Every point on the terrain within the ROI of a transmitter is a potential receiver.

\item[Receiver height:]  $h_r$, the vertical distance between a receiver and its base (the point on the terrain directly below it).  Although this is not conceptually required, for simplicity, all the receivers have the same height, equal to the transmitter height.

\item[Line of sight:] LOS, the straight line between a transmitter and receiver.  The receiver is visible iff the LOS does not intersect the terrain.   This work assumes that the radio wave travels in a straight line, ignoring diffraction and reflection off of the Heaviside layer in the upper atmosphere, 

\item[Viewshed:] a property of a transmitter $T$.  A bitmap recording which of the potential receivers within the ROI of $T$ are visible from $T$.

\item[Visibility index:] a property of a transmitter $T$. The fraction of the potential receivers within the ROI of $T$ that are visible.  In other words, the normalized area of $T$'s viewshed.
    
\end{description}

\begin{figure}[t]
  \centering
  \fbox{\includegraphics[width=.9\linewidth]{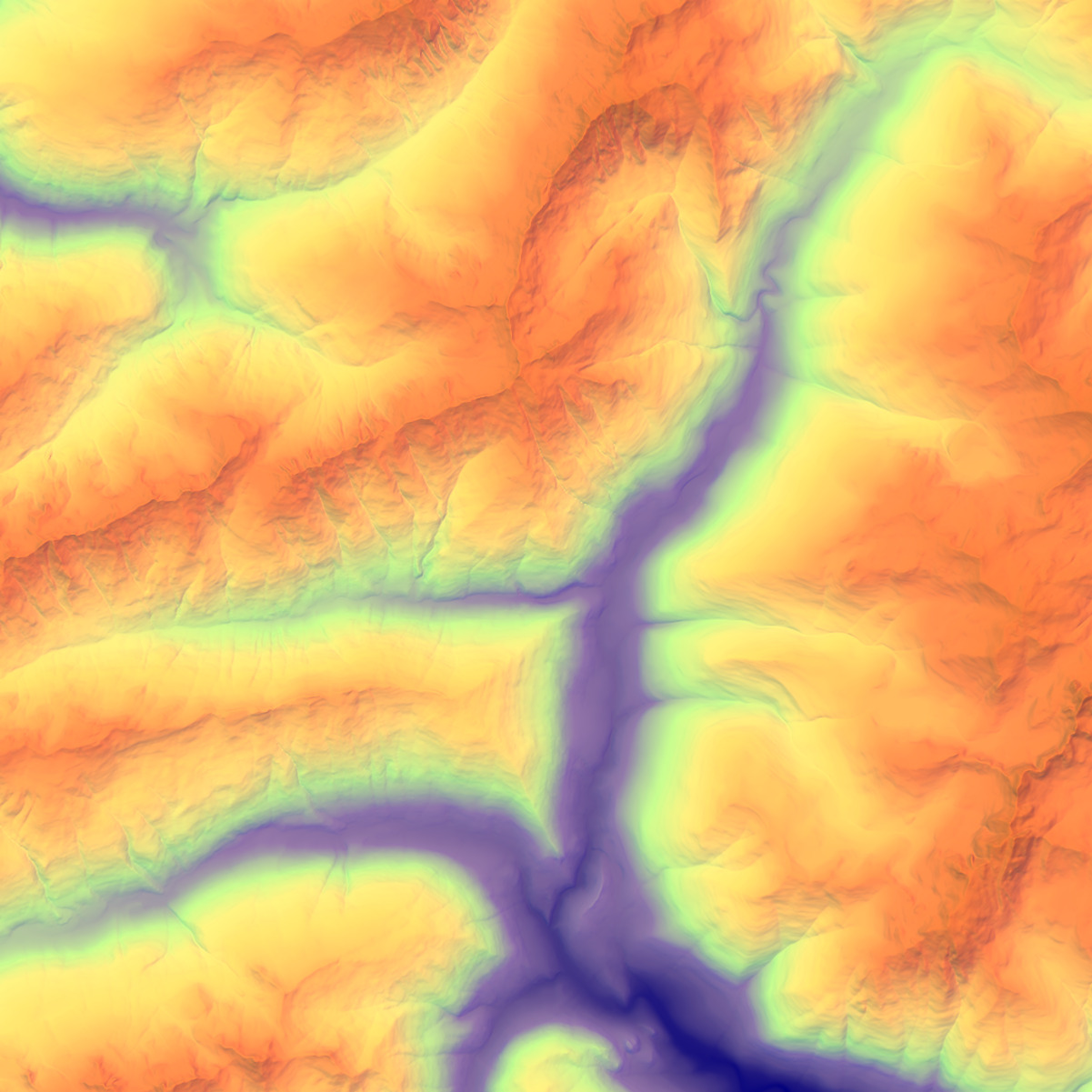}}
  \caption{DEM1000 terrain}
  \Description{DEM1000 terrain}
  \label{f:dem1000-terrain}
\end{figure}

\begin{figure*}[t]
  \centering
  \fbox{\includegraphics[width=1.5in]{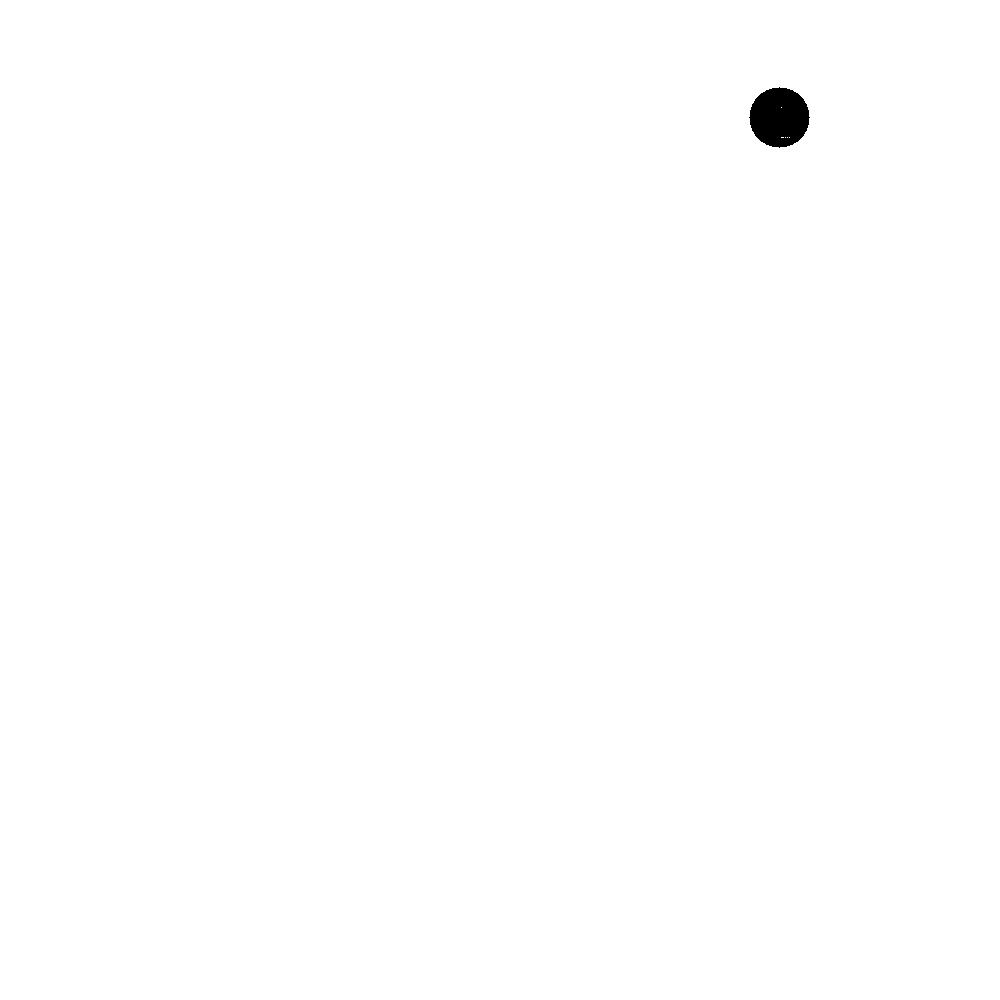}}
  \fbox{\includegraphics[width=1.5in]{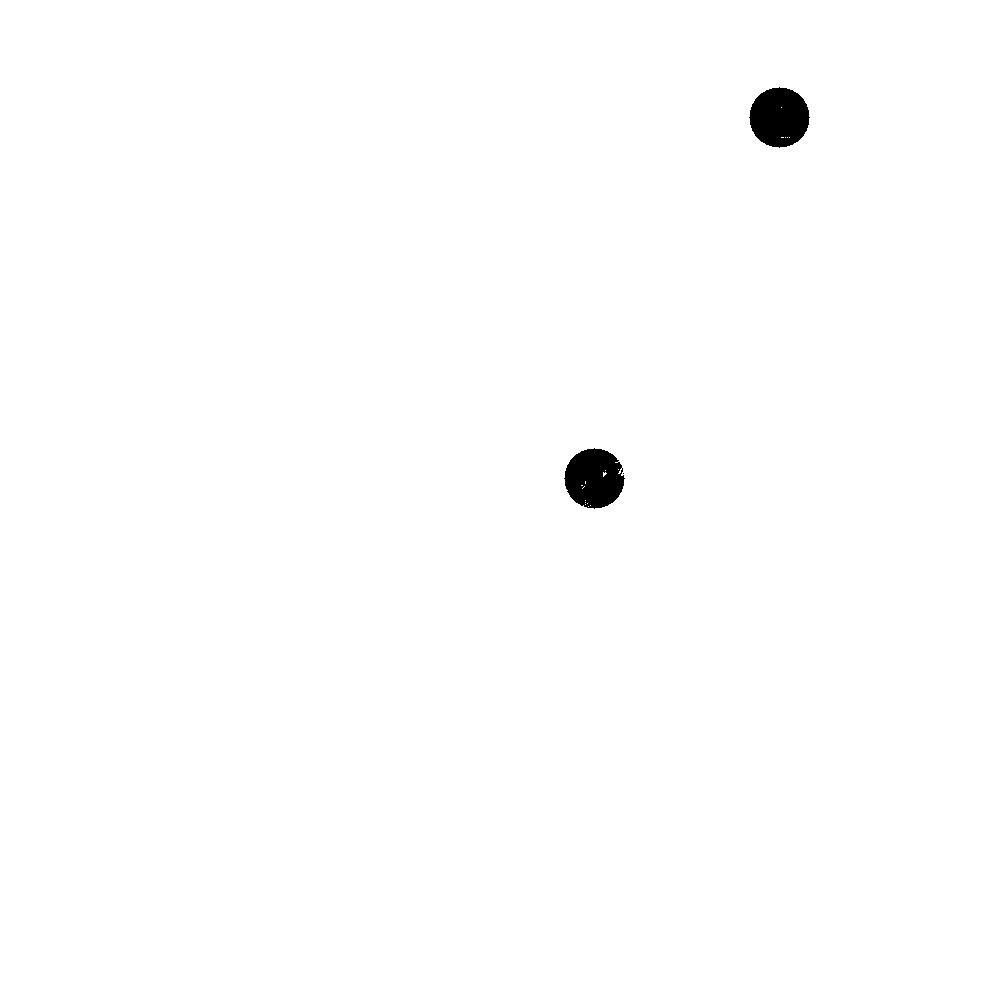}}
  \fbox{\includegraphics[width=1.5in]{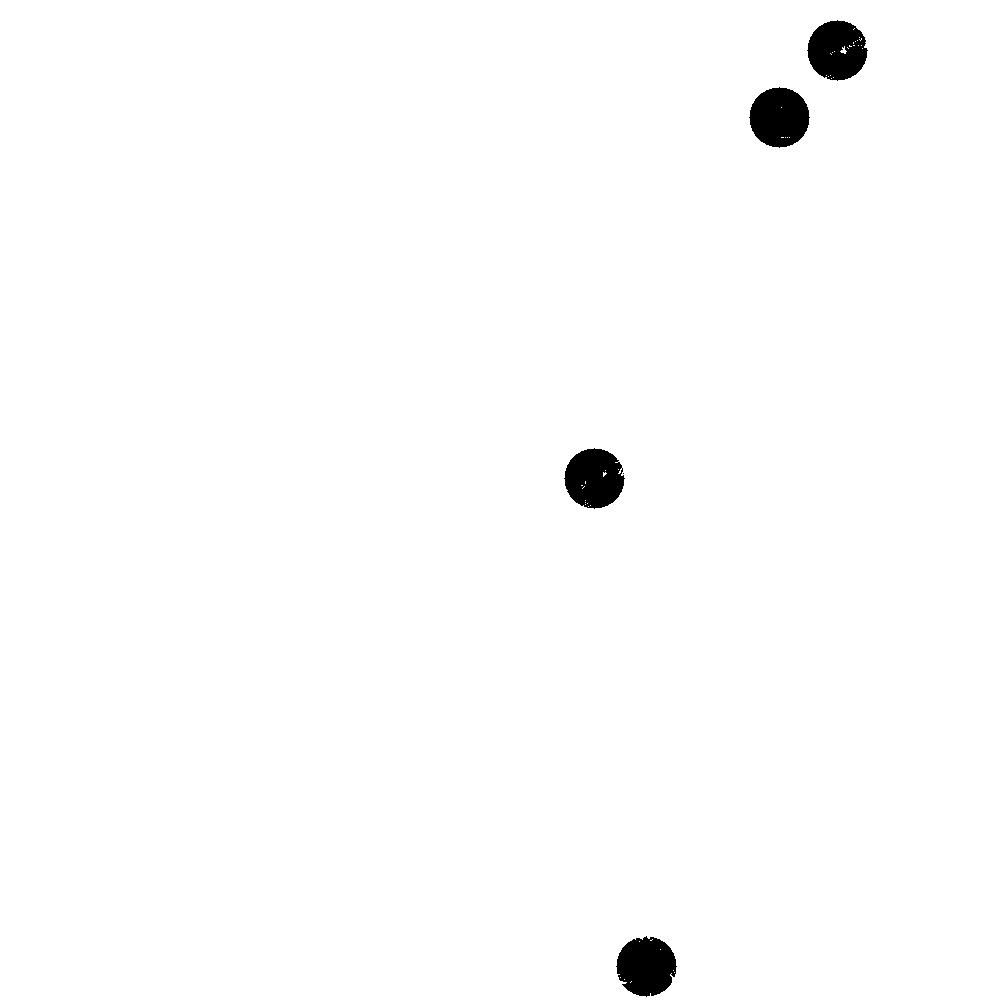}}
  \fbox{\includegraphics[width=1.5in]{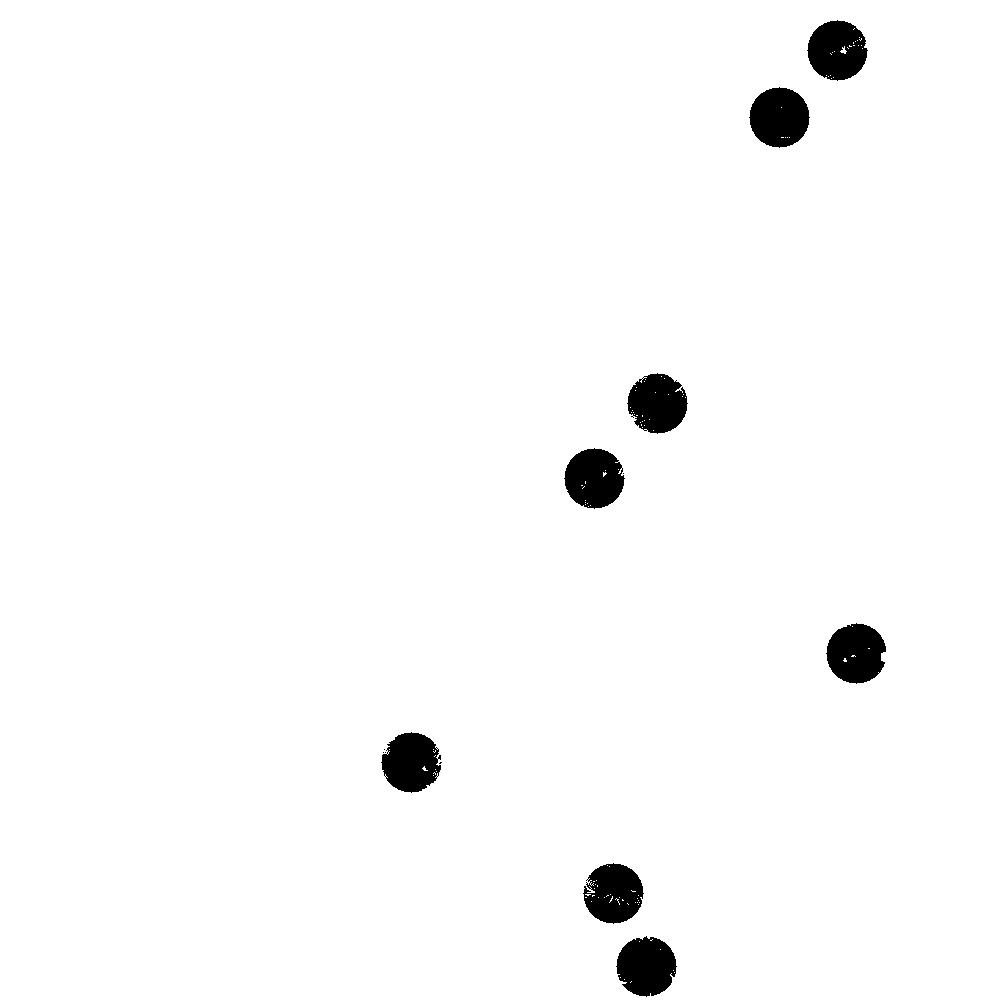}}
  \fbox{\includegraphics[width=1.5in]{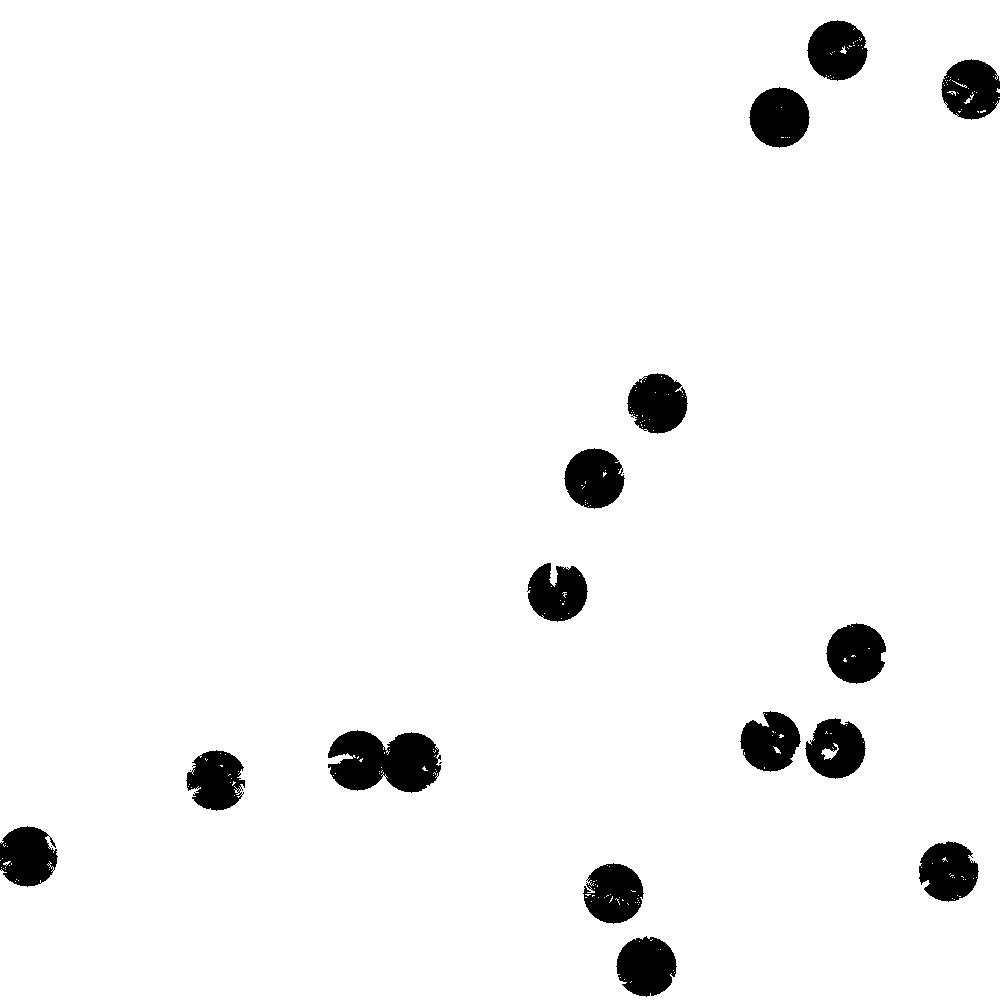}}
  \fbox{\includegraphics[width=1.5in]{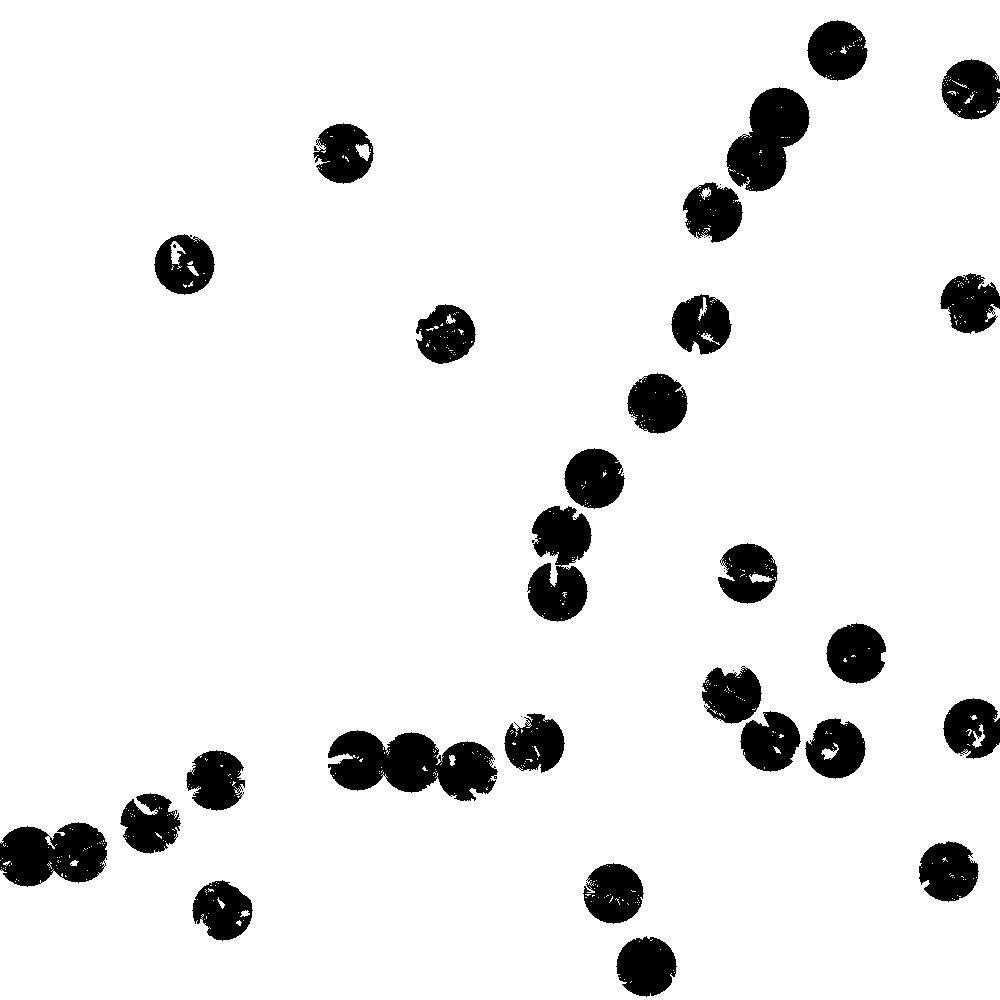}}
  \fbox{\includegraphics[width=1.5in]{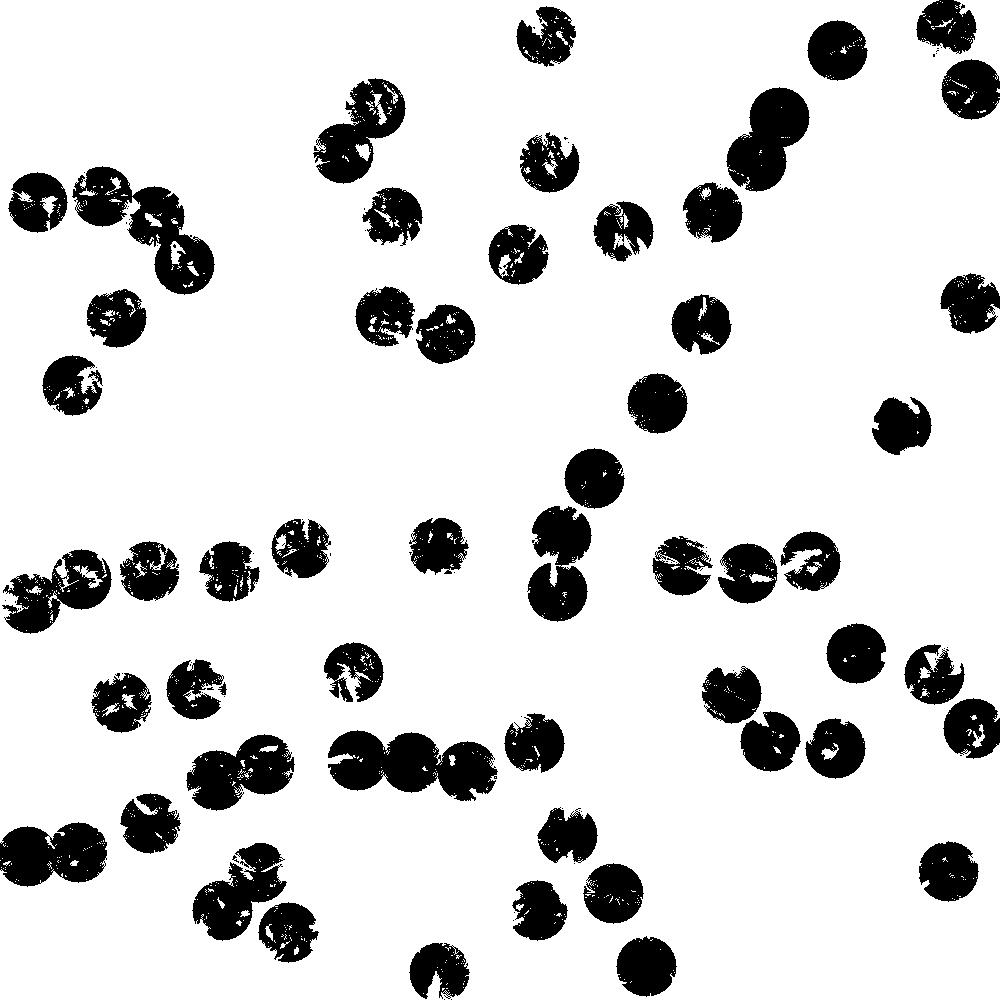}}
  \fbox{\includegraphics[width=1.5in]{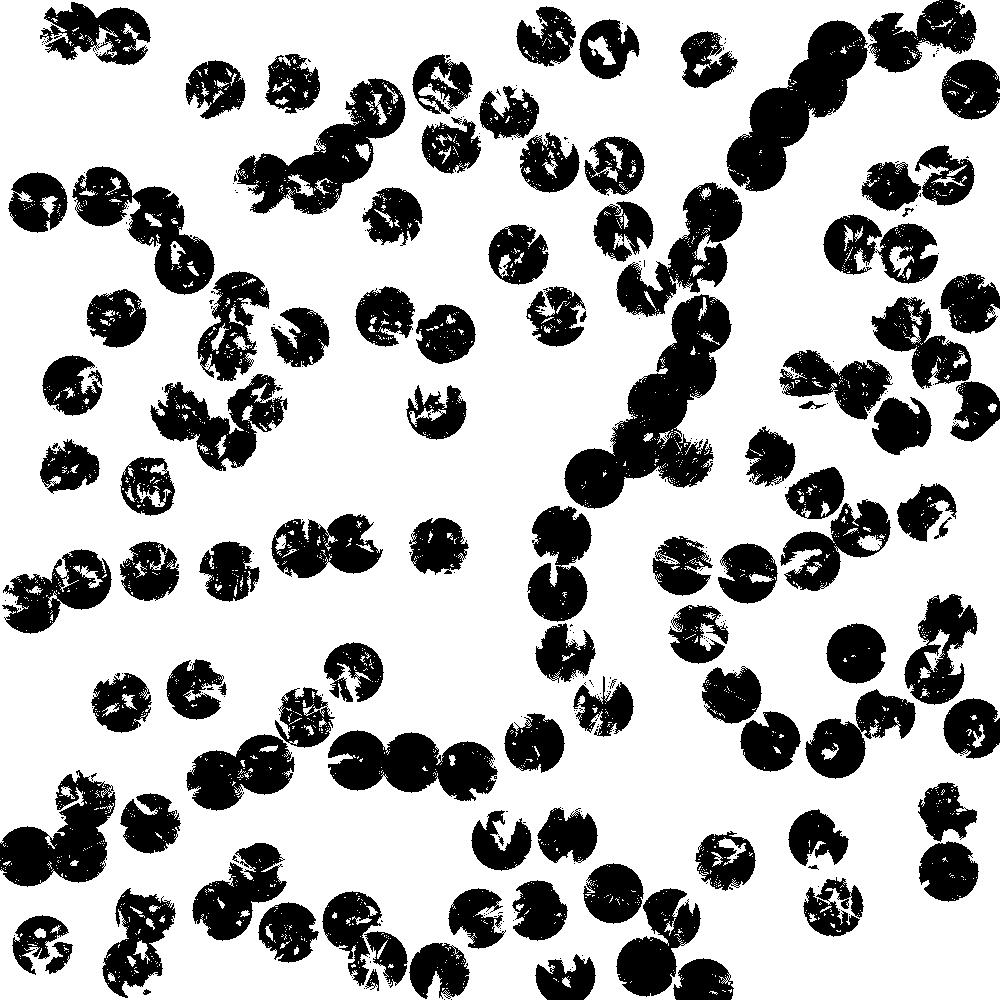}}
  \fbox{\includegraphics[width=1.5in]{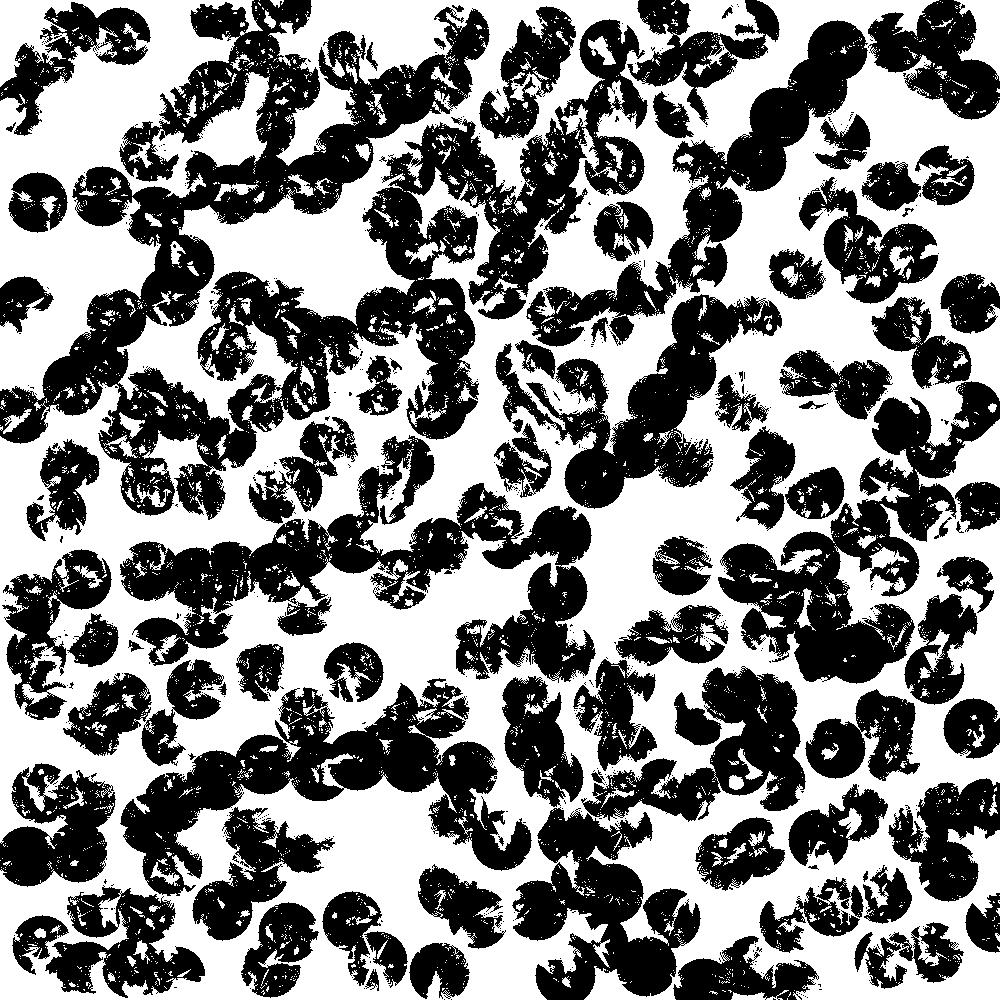}}
  \fbox{\includegraphics[width=1.5in]{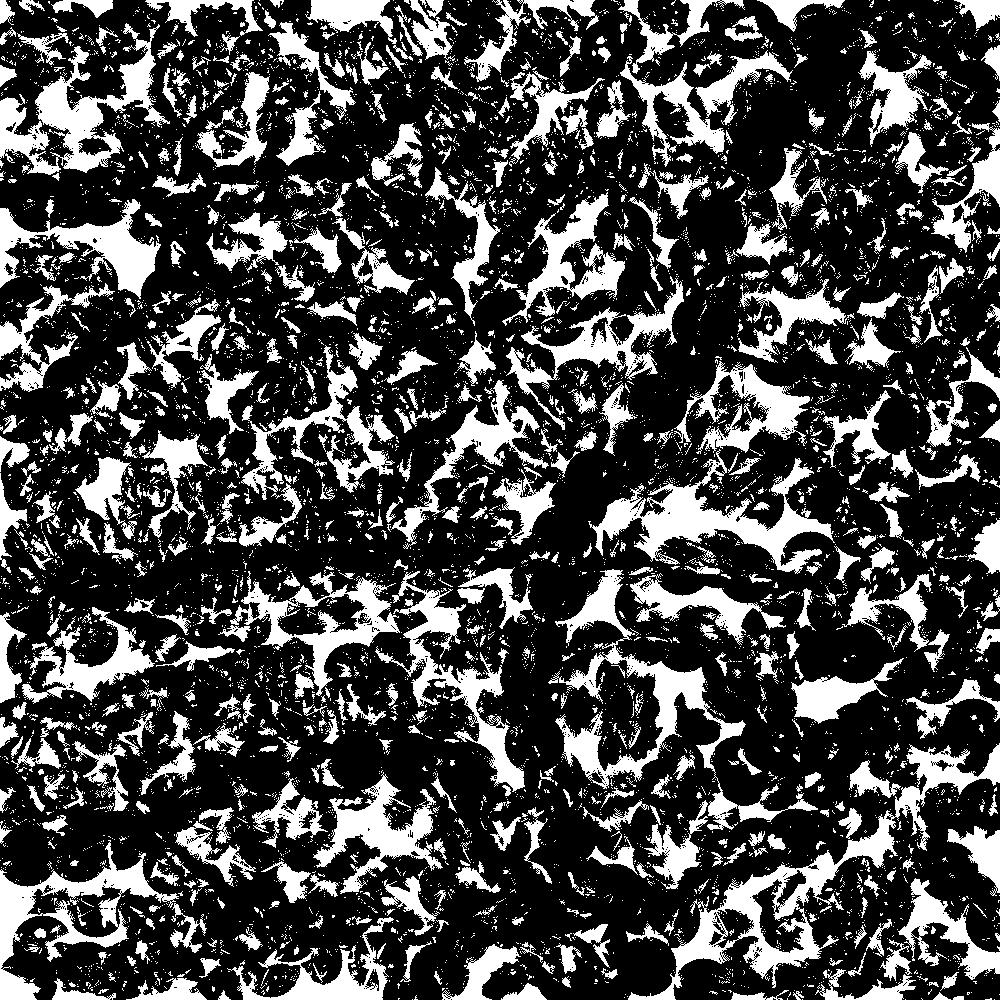}}
  \fbox{\includegraphics[width=1.5in]{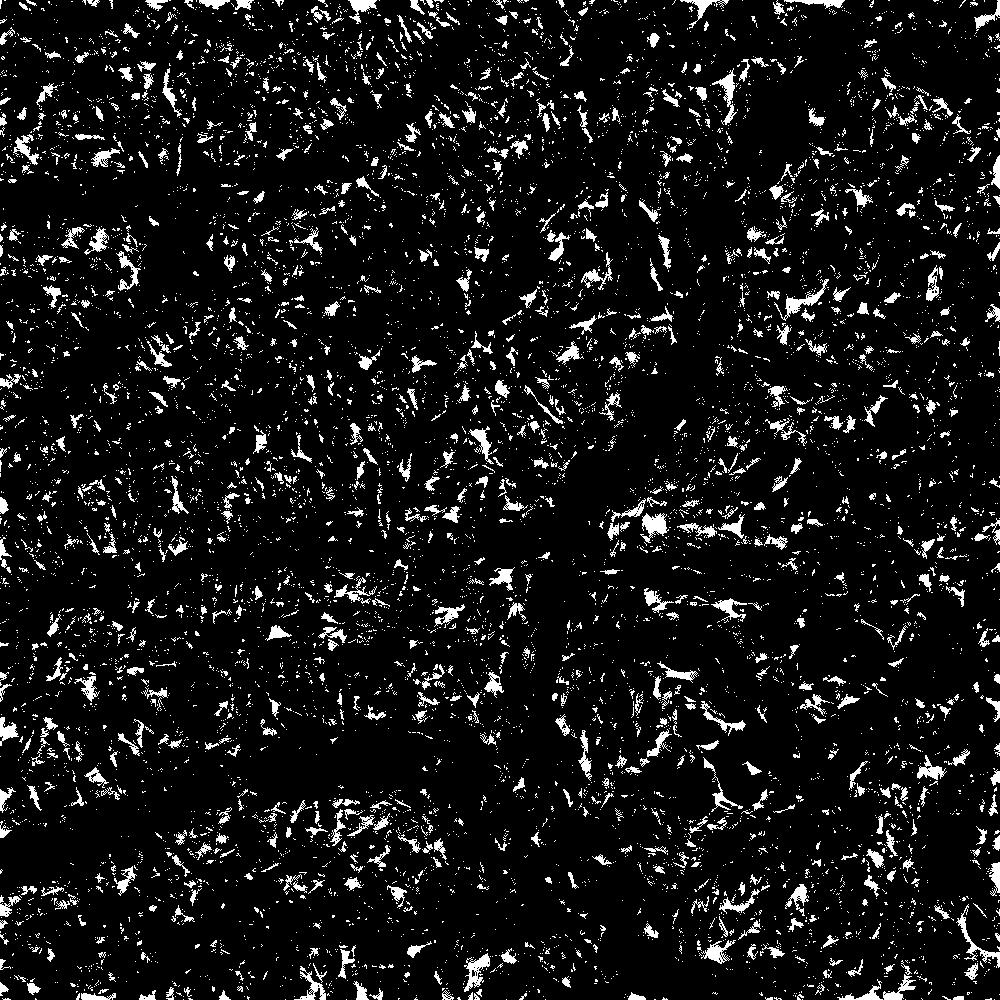}}
  \fbox{\includegraphics[width=1.5in]{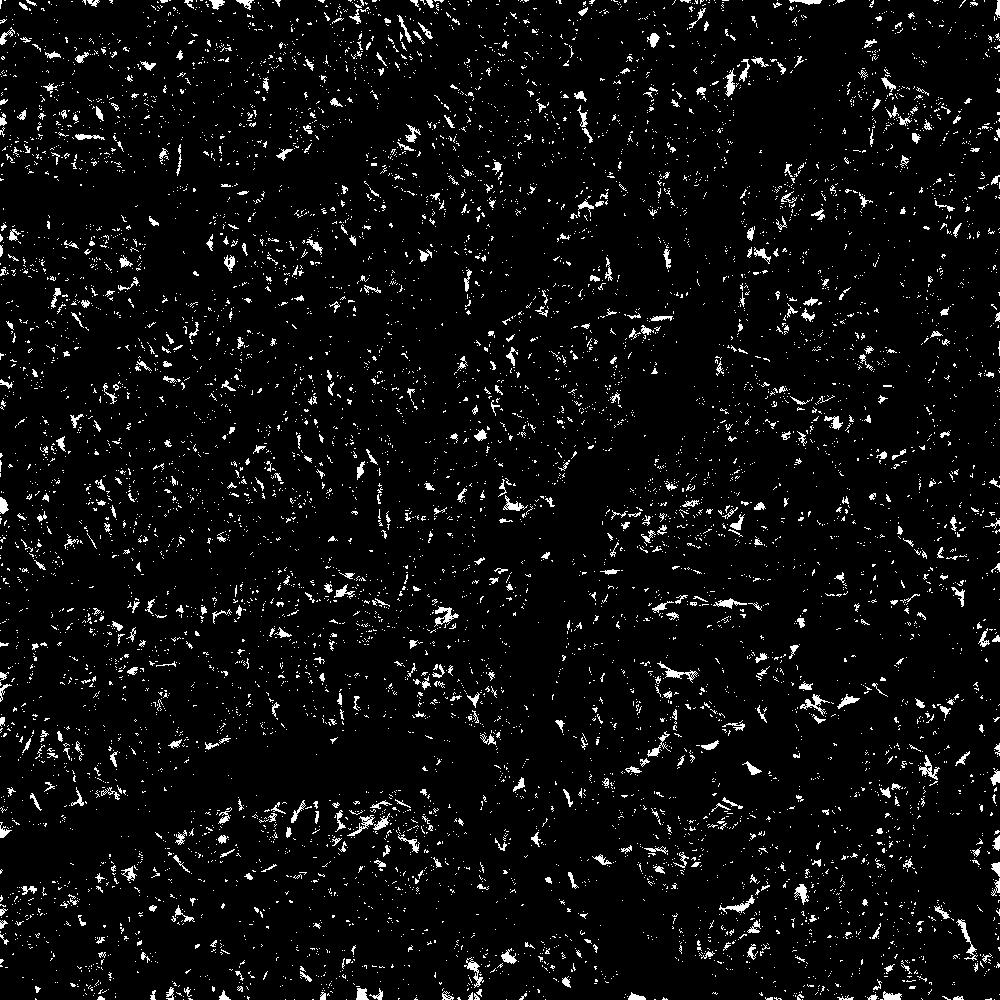}}
  \caption{Cumulative viewsheds for DEM1000 after 1, 2, 4, 8, 16, 32, 64, 128, 256, 512, 1024, and 1264 transmitters sited}
  \Description{Cumulative viewsheds for DEM1000}
\label{f:1000-sheds}
\end{figure*}

\begin{table}
  \caption{DEM1000 test}
  \label{t:1000}
  \begin{tabular}{lr}
    \toprule
Quantity &  Value\\
\midrule
Computer...&\\
.. model &  Xeon E-2276M \\
.. number of cores & 6\\
.. number of hyperthreads & 12\\
.. real memory & 128 GB\\
.. nominal processor speed&2.8 GHz\\
\midrule
    Number of rows & 1000\\
    Number of columns & 1000\\
    Number of elevation posts & 1\,000\,000\\
    Min terrain elevation & 6387\\
    Max terrain elevation & 16344\\
    Transmitter height & 10\\
    Receiver height & 10\\
    Target coverage & 95\% \\
    Radius of interest & 30\\
    Number of blocks the terrain divided into & 100x100\\
    Number of potential transmitters wanted per block & 20\\
    Total number of potential transmitters & 200\,000\\
    Of those, number of transmitters selected & 1264\\
\midrule
    Virtual memory used&142 GB\\
    Real memory used&93 GB\\
    Elapsed time (sec) to ...&\\
    .. read data & 0.025\\
    .. compute estimated visibility indexes & 0.056\\
    .. find potential transmitters & 0.013\\
    .. compute their viewsheds & 1.75\\
    .. find the top transmitters & 2.44\\
    .. in total & 4.30\\
    \bottomrule
  \end{tabular}
\end{table}

\begin{figure*}[t]
  \centering
  \fbox{\includegraphics[width=.9\linewidth]{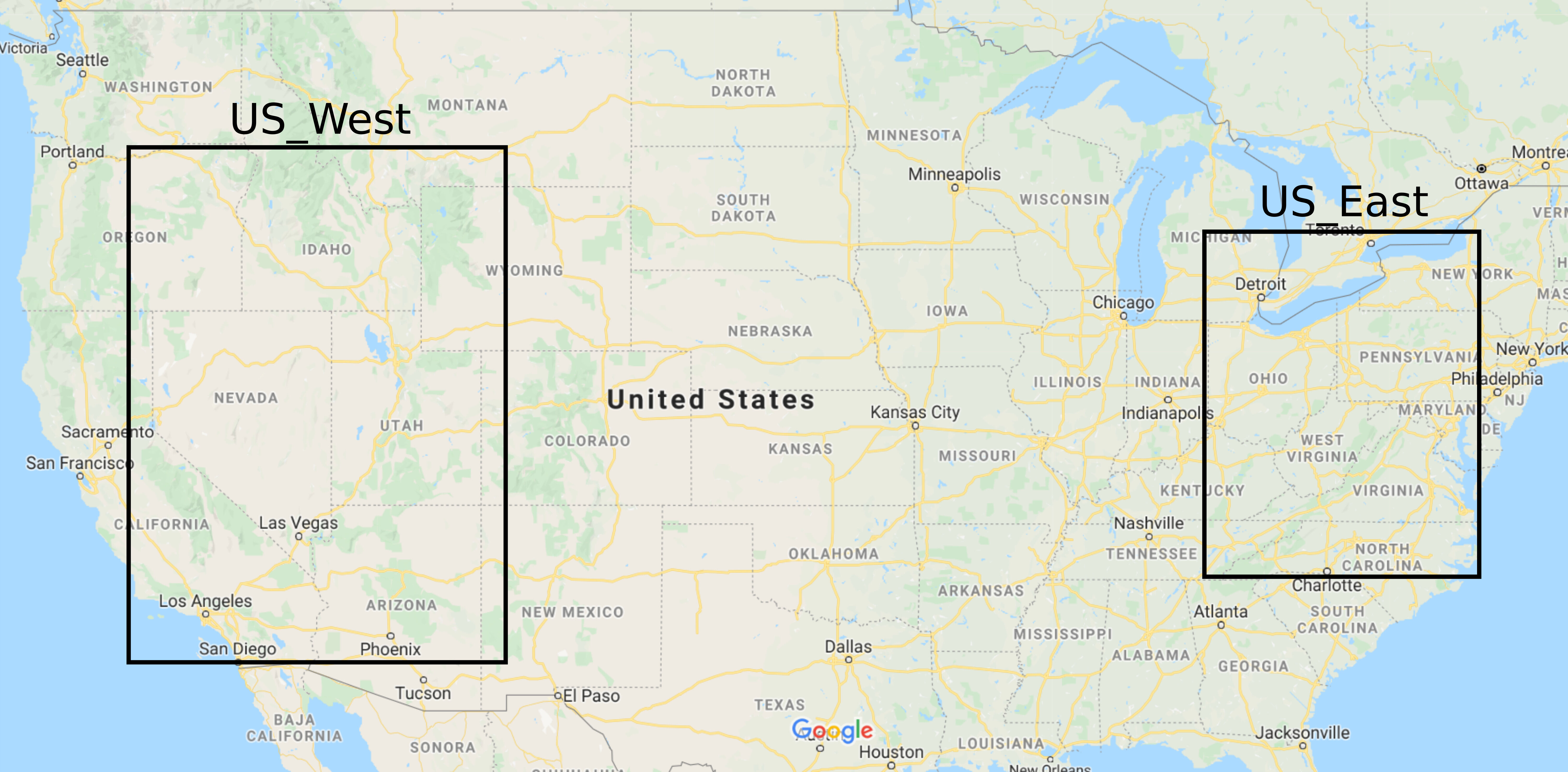}}
  \caption{US West and East dataset locations. Map data ©2020 Google.}
  \Description{US West and East}
  \label{f:usmap}
\end{figure*}

\begin{figure}[t]
  \centering
  \fbox{\includegraphics[width=.9\linewidth]{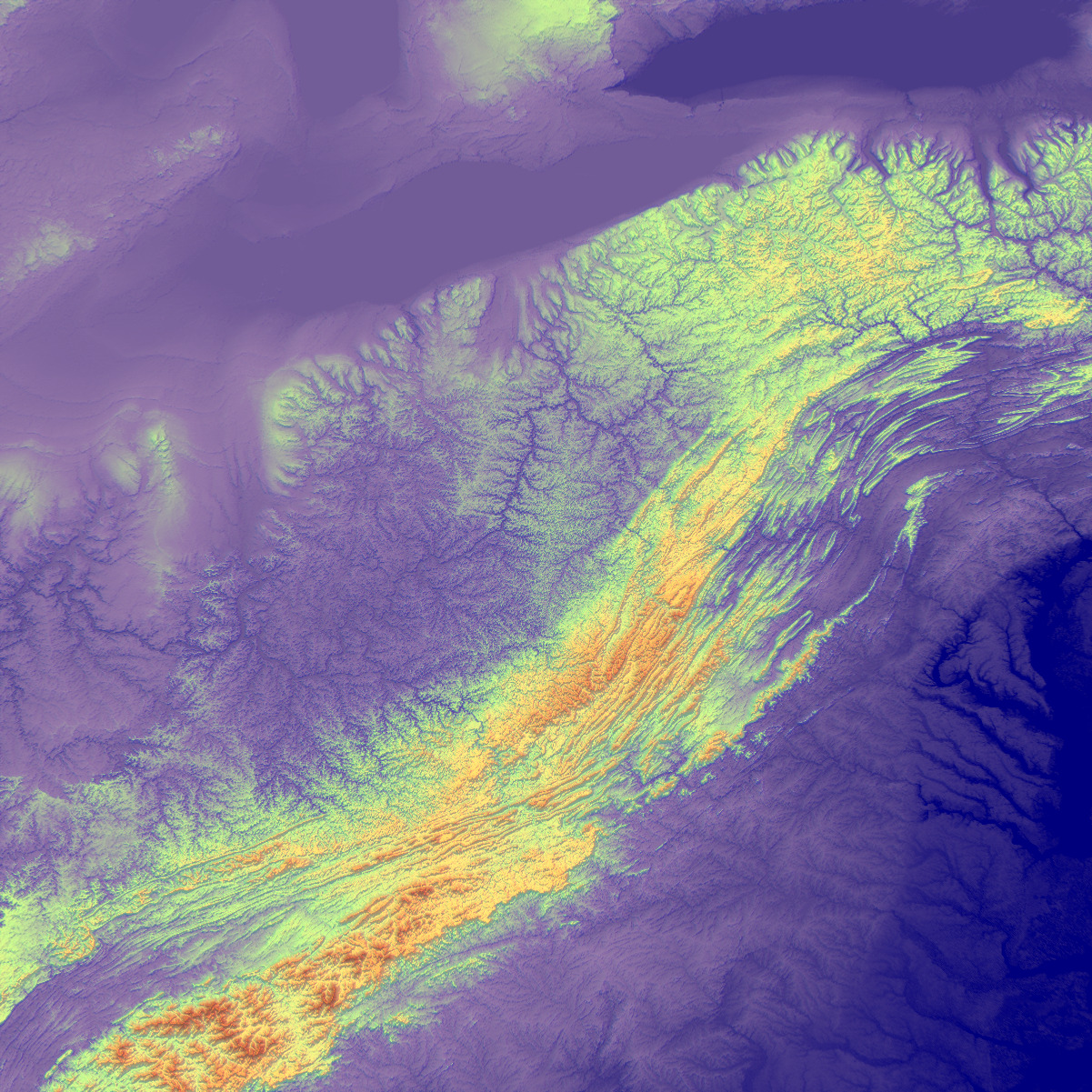}}
  \caption{US East  terrain}
  \Description{US East}
  \label{f:useast}
\end{figure}


\begin{table*}
  \caption{US East tests}
  \label{t:useast}
  \begin{tabular}{lrr}
    \toprule
Quantity & Test 1 value & Test 2 value\\
    \midrule
Computer...&\\
.. model &  \multicolumn{2}{c}{Xeon E-2276M} \\
.. number of cores & \multicolumn{2}{c}{6}\\
.. number of hyperthreads & \multicolumn{2}{c}{12}\\
.. real memory & \multicolumn{2}{c}{128 GB}\\
.. nominal processor speed&\multicolumn{2}{c}{2.8 GHz}\\
    \midrule
    Number of rows & \multicolumn{2}{c}{32000}\\
    Number of columns & \multicolumn{2}{c}{32000}\\
    Number of elevation posts & \multicolumn{2}{c}{1\,024\,000\,000}\\
    Min terrain elevation & \multicolumn{2}{c}{7}\\
    Max terrain elevation & \multicolumn{2}{c}{514}\\
    Transmitter height & \multicolumn{2}{c}{100}\\
    Receiver height & \multicolumn{2}{c}{10}\\
    Target coverage & \multicolumn{2}{c}{95\%} \\
    Radius of interest & 500 & 1000\\
    Number of blocks the terrain divided into & 193x193 & 96x96\\
    Number of potential transmitters wanted per block & 20 & 20\\
    Total number of potential transmitters & 744\,980 & 184\,320\\
    Of those, number of transmitters selected & 6543 & 5000\\
    \midrule
    Elapsed time (sec) to ...&\\
    .. read data & 24 & 22\\
    .. compute estimated visibility indexes & 149 & 188\\
    .. find potential transmitters & 14 & 14\\
    .. compute their viewsheds & 2145 & 2523\\
    .. find the top transmitters & 1501 & 1984\\
    .. in total & 3834 & 4732\\
    \bottomrule
  \end{tabular}
\end{table*}

\section{Multiple transmitter siting}

How should we best site (i.e., determine locations for) a set of radio transmitters $t_i$, to cover some terrain, so that the maximum number of receivers, $r_j$ can be accessed, or in other words, are visible?


The most important current application of this problem is in siting cell phone towers, and so this paper uses that terminology --- \emph{transmitters, receivers}, etc.  However this problem is a few decades old, originally being of interest in  the surveillance and environental visual domain.   They use different a terminology of \emph{observers} and \emph{targets}.  There we might have been siting a set of observers so that they could jointly see the most terrain.  We even have wanted that the unsurveilled terrain consist of small separated regions instead of large connected regions that a smuggler might use.  Mathematically, these are the same problem with different words.

\section{Terrain representation}

A formally grounded study of this problem would need a model for terrain.  However, this important, and difficult, problem is not totally solved.  It is hard because terrain has unusual properties.

\begin{enumerate}

\item \textbf{Up} and \textbf{down} are different for terrain.  There are many sharp local maxima (peaks), but only few local minima (endorheic lakes), and they are broad, not sharp.

\item There are \textbf{long-range monotonic features}, aka river systems.

\item The many mostly smooth regions are interspersed with occasional \textbf{discontinuities}, aka cliffs.

\end{enumerate}

This is important because those properties are not a good match for standard mathematical representations like Fourier series.    In other engineering domains, such as signal processing, a function, perhaps the Fourier expansion
\begin{equation*}
  \sum_{k=0}^N a_k \cos kt + \sum_{k=1}^N b_k \sin kt
\end{equation*}
\noindent might be fitted to a sequence of sample points, and the physics of the problem will tend to match the math.  That is, the mathematical operation of truncating the series at some $N$ to smooth out small features aligns with the physical operation of lo-pass filtering images or audio signals.  This match does not apply to terrain.

Such a lo-pass filter would remove discontinuities like cliffs, which are, for many applications, the most important features of the terrain.  Cliffs are visually recognizable, and affect mobility and drainage.    The triangulated irregular triangle (TIN) representation also has this limitation.

Therefore, this paper will represent terrain with an equally spaced array of elevation posts, or a Digital Elevation Model.   The DEM has its own limitations, but at least the representation is simple, and parallelization of the code is easier.  ``Equally spaced'' is not possible over large regions.   A bigger problem is what the elevation number at the post means.  Here are some possibilities.

\begin{enumerate}
\item The reported elevation might be the terrain elevation at that precise point, to the extent possible.   If the ideal terrain is $z(x,y)$ for real numbers $x$ and $y$, then $z_{ij} = z(x_i, y_j)$. 

\item It might be a convolution or average over a region such as the region halfway to the next post.  E.g.,
\begin{equation*}
  z_{ij} = \int_{x_i-1/2}^{x_i+1/2} \int_{y_j-1/2}^{y_j+1/2} z(x,y) dx dy  \ \ \ \ .
\end{equation*}
\noindent A \emph{sinc} function would be better than the above simple average since sinc goes to zero gradually instead of dropping off sharply.
  
\item The reported elevation might be the max elevation over the region, or some other function chosen to be  useful to the desired application.
\end{enumerate}

At this point, we have only the elevation array, and have no more information about the real terrain.   However, we may need elevations at points between the elevation points.   So we need an algorithm  to interpolate elevations between adjacent posts.  The particular problem here is deciding whether the terrain blocks a line of sight passing between adjacent two posts.   
There is no one best algorithm, since different applications have different needs.    Isolated high elevations are of great interest to aviators.  Cliffs affect land mobility.   Monotonicity affects hydrography.


\section{Terrain visibility}

The terrain will be represented as an array of elevation posts $z_{ij}$.   $i$ and $j$ can be considered to be $x$ and $y$ coordinates, respectively, if the elevation posts are 1 apart.  We must determine whether transmitter $T$, whose 2D base  is $(t_x, t_y)$, and whose 3D location is   $(t_x, t_y, h_t + z_{t_x, t_y})$ can see the receiver $R$, 
whose 2D base is $(r_x, r_y)$, and whose 3D location is $(r_x, r_y, h_r + z_{r_x, r_y})$.
This requires determining if a straight line, the LOS, drawn from $(t_x, t_y, h_t + z_{t_x, t_y})$ to $(r_x, r_y, h_r + z_{r_x, r_y})$  intersects the terrain.    In general, the LOS runs between adjacent pairs of elevation posts, so we must interpolate elevations, in this case with a linear interpolation.

\section{Prior art}
Ray\cite{ray-phd} and Franklin and Ray\cite{fr-hinbv-94} described several fast programs to compute viewsheds and weighted visibility indices for observation points in a raster terrain. These programs explore various tradeoffs between speed and accuracy. They analyzed many cells of data; there is no strong correlation between a point’s elevation and its weighted visibility index. However, the, very few, high visibility points tend to characterize features of the terrain.  Franklin\cite{wrf-site} presented an experimental study of a new algorithm that synthesizes separate programs, for fast viewshed, and for fast approximate visibility index determination, into a working testbed for siting multiple transmitters jointly to cover terrain from a full level-1 DEM, and to do it so quickly that multiple experiments are easily possible.  Franklin and Vogt \cite{wrf-cv-siting-isprs, wrf-siting-apr2004,wrf-sdh2006} described two projects for siting multiple transmitters on terrain.  Vogt\cite{vogt-ms} studied the effect of varying the resolution.

A variation of this problem has recently been employed for siting a fixed number of terrestrial laser scanners on a terrain, ~\citet{SitingLaserScanners}. The authors employed a Simulated Annealing heuristic in their method, but focused only on very small instances with up to 6 transmitters on a $450\times450$ terrain.

Tracy et al\cite{dt-wrf-spie-2007}, Tracy\cite{tracy-phd}, and Franklin et al\cite{acmgis07} extended multiple transmitter siting to compute smugglers paths to avoid the transmitters.

Andrade et al\cite{andrade-geoinfo-ext-viewshed-2008} presented an external memory viewshed program, which managed paging the data better than the virtual memory manager (because it understood the data access pattern better).  Magalhães et al\cite{magalhaes-ijcisim-2011} and Ferreira et al\cite{ferreira-acmgis-2012,chaulio-geoinfo-2013,chaulio-jidm-2014,chaulio-tiledvs-tsas-2016} improved the external memory algorithm and also presented a parallel viewshed algorithm in external memory.  Pena et al\cite{iceis-2014,bigspatial-2014}, Li\cite{wenli-phd} and Li et al\cite{li-acmgis-2014,wenli-gpu-siting-2016} presented parallel observer siting algorithms running on GPUs.

It is also possible to consider receivers that have a certain quality, or are visible with some given probability, Akbarzadeh et al
\citep{Akbarzadeh:2013:PSM}.  We might  add constraints such as intervisibility, where transmitters are required to be visible from other transmitters.  The transmitters and receivers might be mobile, Efrat \citep{Efrat:2012:EAP}.  Placing transmitters at different positions might have different costs.

The Modeling and Simulation community, which is disjoint from this community, discusses line-of-sight (with comparisons of various LOS algorithms) in \citet{teclos}, and the relation of visibility to topographic features,  \citet{jl:vdtfdem}.  \citet{n-tv-94,champion2002} studied line-of-sight on natural terrain defined by an $L_1$-spline.

The parallelization of line-of-sight and viewshed algorithms on terrains using GPGPU or multi-core CPUs is an active topic.  \citet{Strnad:2011:PTV} parallelized the line-of-sight calculations between two sets of points---a source set and a destination set---on a GPU, and implemented it on a multi-core CPU for comparison.  \citet{Zhao:2013:PCA} parallelized Franklin's R3 algorithm \citep{Franklin:1994:HNB} to compute viewsheds on a GPU.  The parallel algorithm combines coarse-scale and fine-scale domain decompositions to deal with memory limit and enhance memory access performance.  \citet{Osterman:2012:IRM} parallelized the \emph{r.los} module (R3 algorithm) of the open-source GRASS GIS on a GPU.  \citet{Osterman:2014:IPI} also parallelized Franklin's R2 algorithm \citep{Franklin:1994:HNB}.  \citet{Axell:2015:CBG} parallelized and compared the R2 algorithm on a GPU and on a multi-core CPU.  \citet{Bravo:2015:EIF} parallelized Franklin's XDRAW algorithm \citep{Franklin:1994:HNB} to compute viewsheds on a multi-core CPU, after improving its IO efficiency and compatibility with SIMD instructions.  \citet{Ferreira:2014:PAV, chaulio-tiledvs-tsas-2016} parallelized the sweep-line algorithm of \citet{VanKreveld:1996:VSA} to compute viewsheds on multi-core CPUs.
Qarah and Tu \cite{qarah2019fast} presented a fast GPU sweep-line viewshed algorithm, while Jianbo et al\cite{jianbo2019parallel} used Spark.  Wu et al\cite{wu2019hivewshed} presented an interactive online multiple transmitter viewshed analysis system.  

\citet{Rana:2003:FAV} proposed using topographic feature points, instead of random points, as receivers when estimating visibility indices.  \citet{Wang:2000:GVW} proposed a viewshed algorithm that uses a plane instead of lines of sight in each of 8 standard sectors around the transmitter to approximate the local horizon.  The algorithm is faster but less accurate than XDRAW.  \citet{Israelevitz:2003:FAA} extended XDRAW to increase accuracy by sacrificing speed.
Wang and Dou\cite{wang2020fast} showed fast algorithm for filtering possible viewpoints.  Eli{\c{s}}\cite{elics2017terrain} studied using multiple guard towers on terrain.   Zhu et al\cite{zhu2019hixdraw} improved XDRAW to remote chunk distortion.  Lin et al\cite{lin2018improved} studied intervisibility.

Gillings\cite{gillings2015mapping} used viewshed analysis in archeology.  Shi and Xue\cite{shi2016deriving} also minimized the number of transmitters while maximizing coverage.   Prescott and Toma\cite{prescott2018multiresolution} used a multiresolution approach.  Yu et al\cite{yu2017synthetic} used a synthetic visual plane technique.  Shrestha and Panday\cite{shrestha2017faster} improved on  R3.  Baek and Choi\cite{baek2018comparison} compared different viewshed algorithms, using factors such as a 3D Fresnel zone.    Efrat et al\cite{Efrat:2012:EAP} used visibility to pursue moving evaders.

\begin{figure}[t]
  \centering
  \fbox{\includegraphics[width=.9\linewidth]{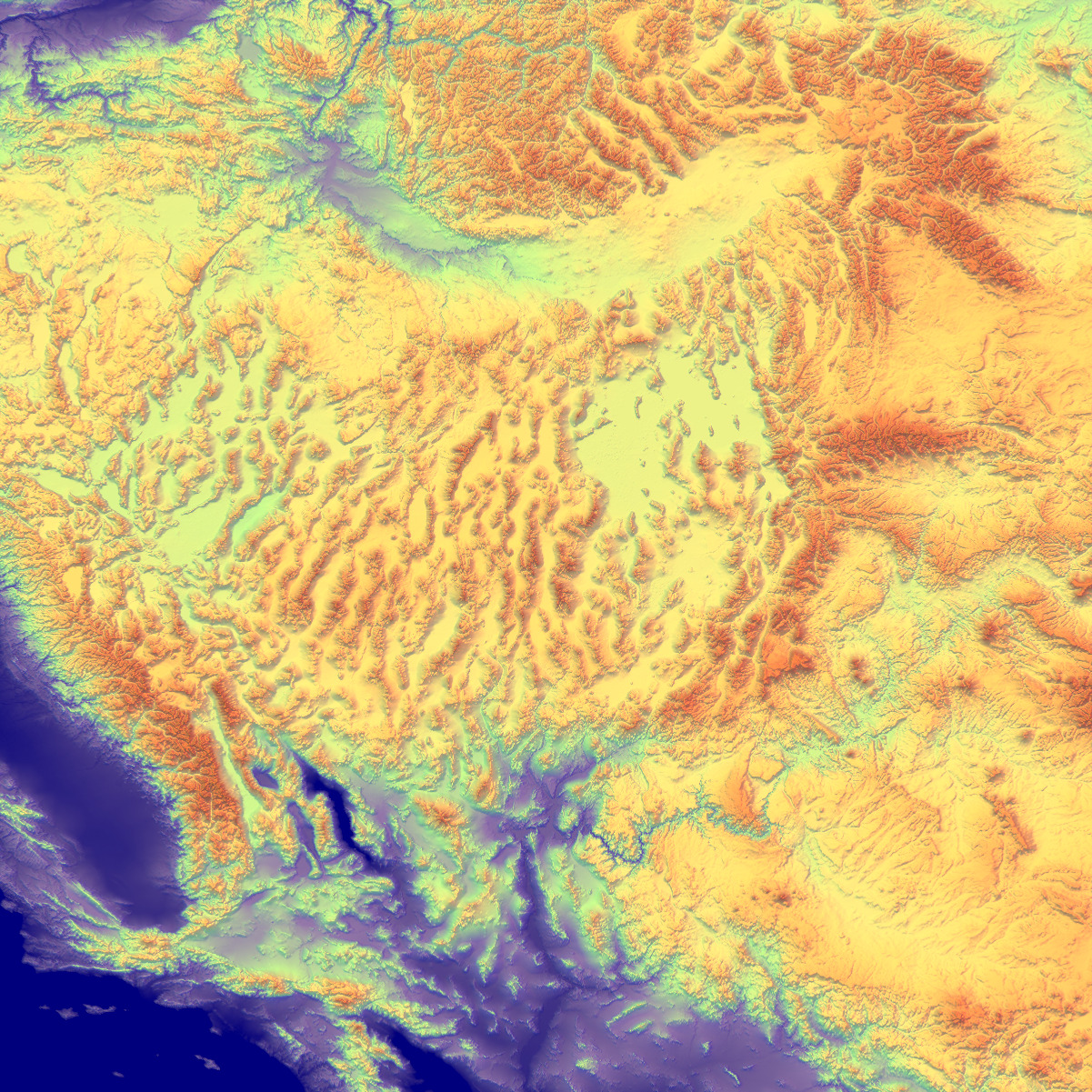}}
  \caption{US West terrain.}
  \Description{US West}
  \label{f:uswest}
\end{figure}

\begin{table*}
  \caption{US West tests}
  \label{t:uswest}
  \begin{tabular}{lrr}
    \toprule
Quantity & Test 1 value&Test 2 value\\
    \midrule
Computer...&\\
.. model &  \multicolumn{2}{c}{Xeon  E5-2660 v4} \\
.. number of cores & \multicolumn{2}{c}{14}\\
.. number of hyperthreads & \multicolumn{2}{c}{28}\\
.. real memory & \multicolumn{2}{c}{256 GB}\\
.. nominal processor speed&\multicolumn{2}{c}{2 GHz}\\
    \midrule
    Number of rows & \multicolumn{2}{c}{46400}\\
    Number of columns & \multicolumn{2}{c}{46400}\\
    Number of elevation posts & \multicolumn{2}{c}{2\,152\,960\,000}\\
    Min terrain elevation & \multicolumn{2}{c}{80}\\
    Max terrain elevation & \multicolumn{2}{c}{2786}\\
    Transmitter height & \multicolumn{2}{c}{100}\\
    Receiver height & \multicolumn{2}{c}{10}\\
    Target coverage & \multicolumn{2}{c}{95\%} \\
    Radius of interest & 1000 & 2000\\
    Number of blocks the terrain divided into & 139x139 & 70x70\\
    Number of potential transmitters wanted per block & 20 & 20\\
    Total number of potential transmitters & 386420 & 98000\\
    Of those, number of transmitters selected & 5647 & 3347\\
    \midrule
    Virtual memory used&\multicolumn{2}{c}{195 GB}\\
    Real memory used&\multicolumn{2}{c}{194 GB}\\
    Elapsed time (sec) to ...&\\
    .. read data & 118 & 109\\
    .. compute estimated visibility indexes & 130 & 143\\
    .. find potential transmitters & 9 & 8\\
    .. compute their viewsheds & 1706 & 2132\\
    .. find the top transmitters & 3510 & 3116\\
    .. in total & 5473 & 5509\\
    CPU parallelism & \multicolumn{2}{c}{32x}\\
    \bottomrule
  \end{tabular}
\end{table*}

\begin{figure*}[t]
  \centering
  \fbox{\includegraphics[width=1.5in]{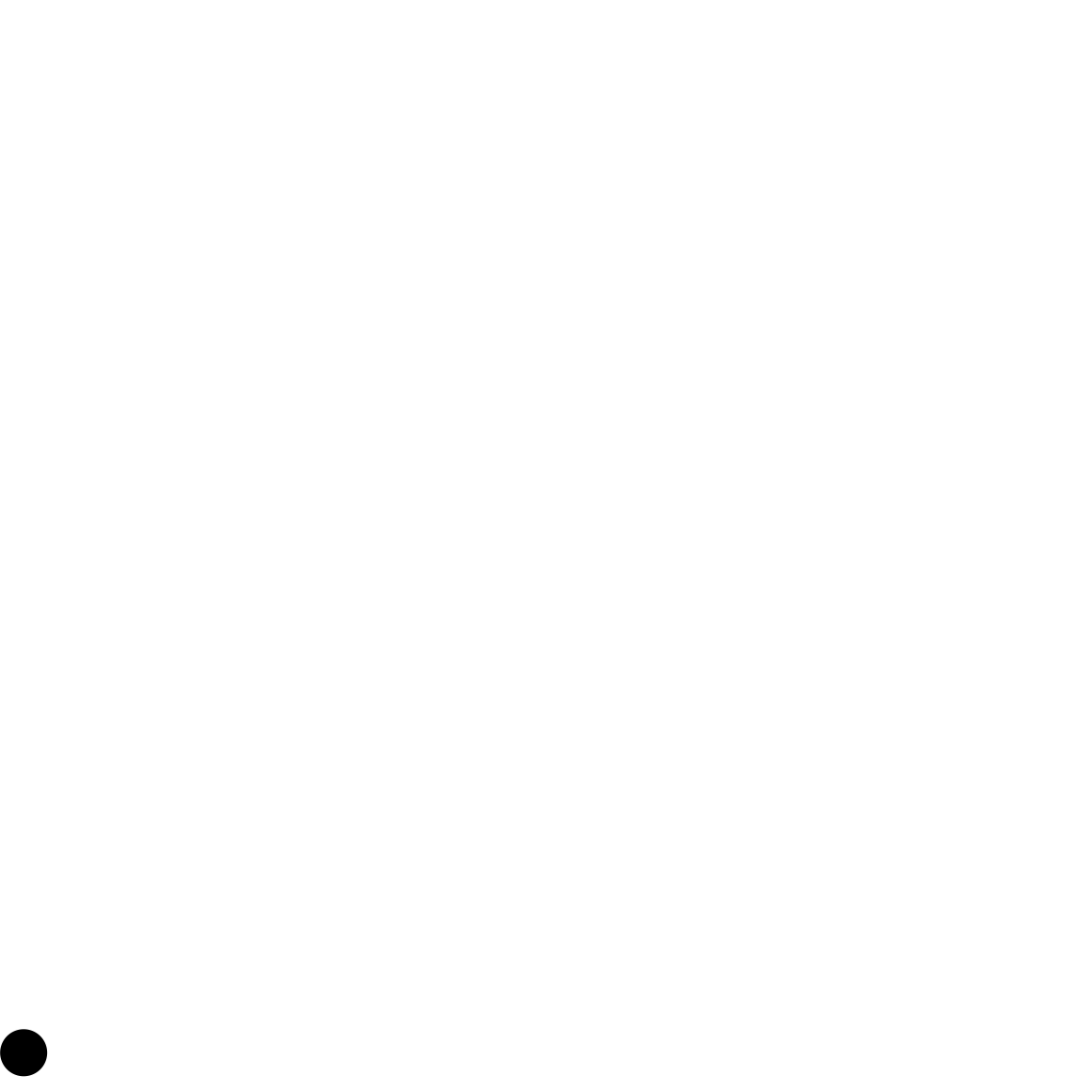}}
  \fbox{\includegraphics[width=1.5in]{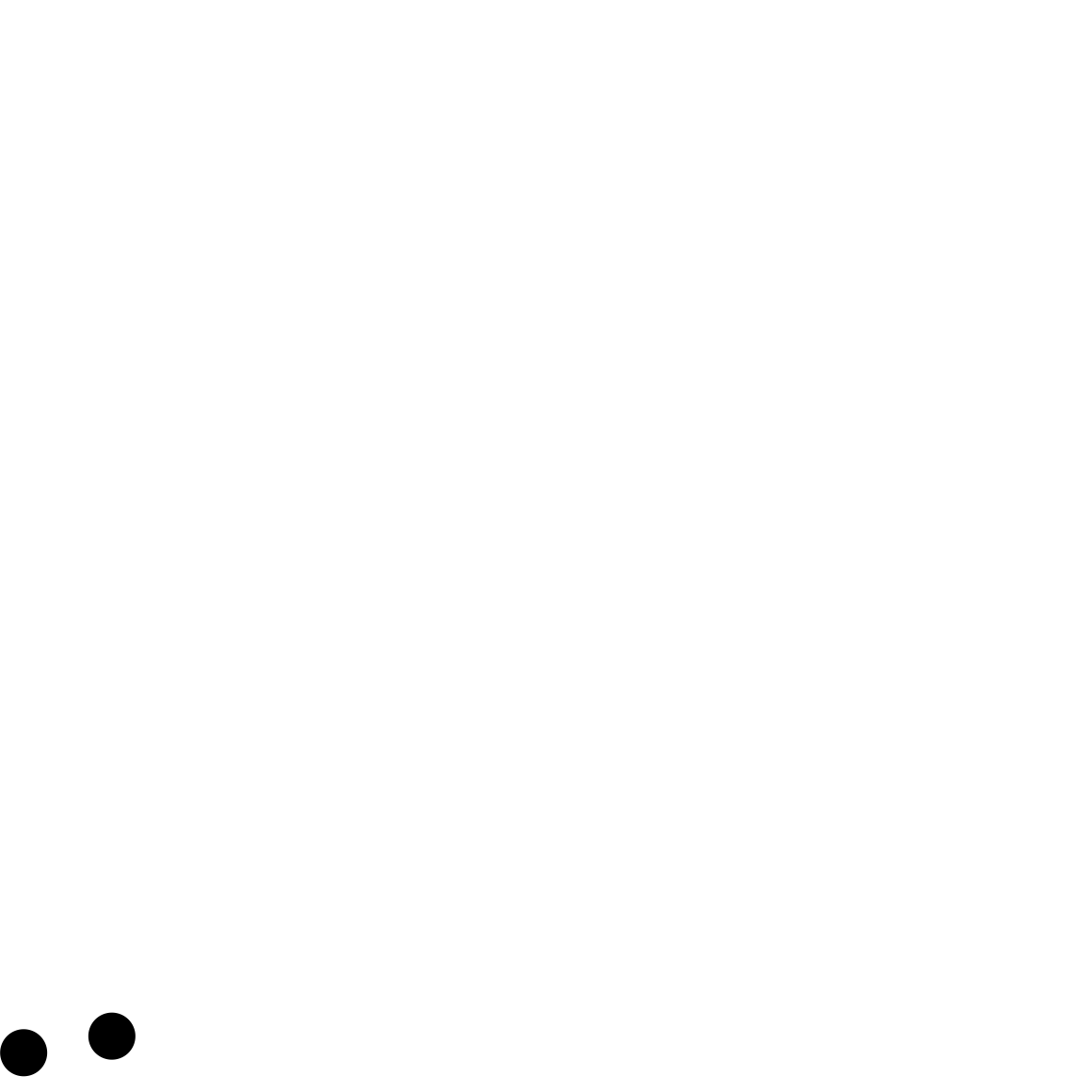}}
  \fbox{\includegraphics[width=1.5in]{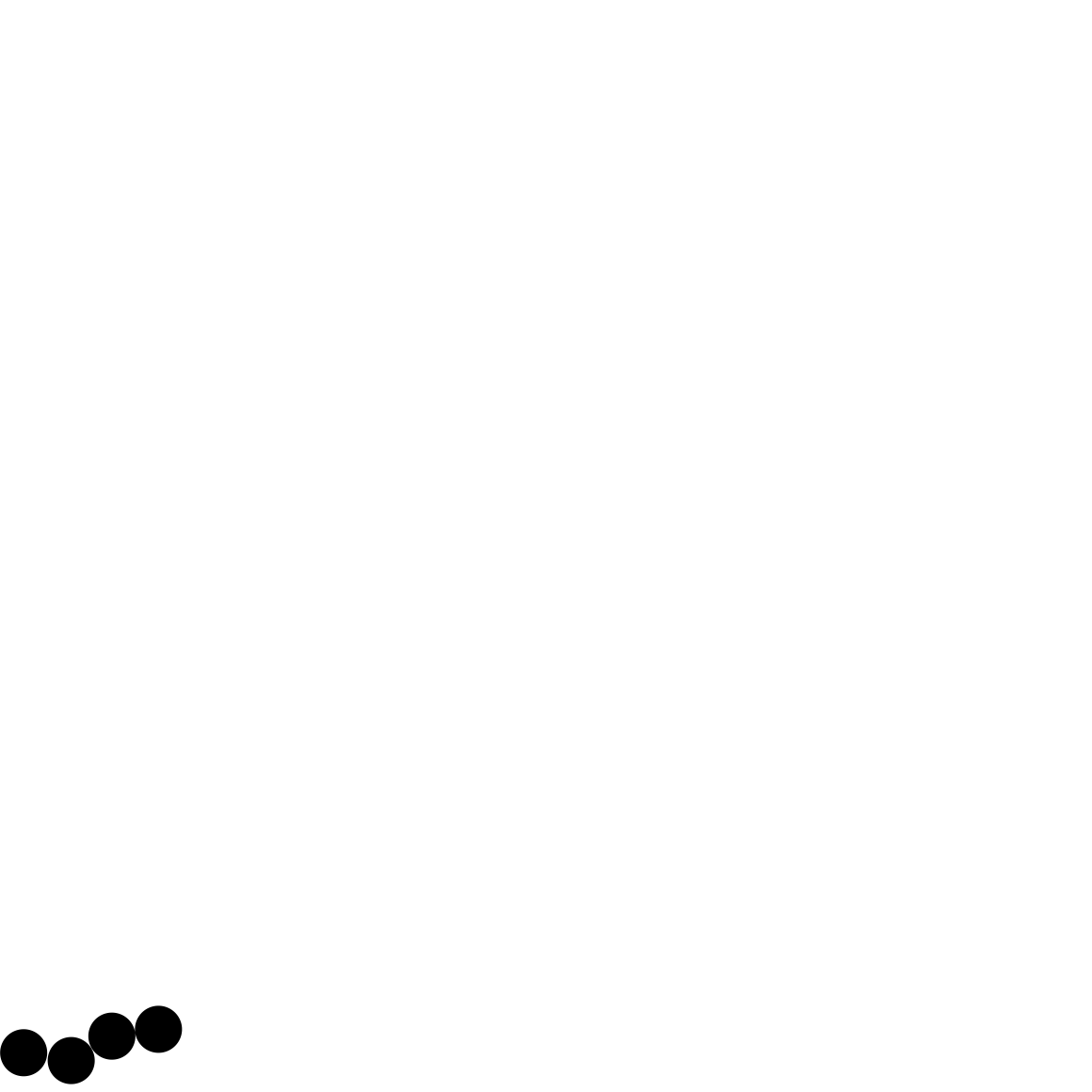}}
  \fbox{\includegraphics[width=1.5in]{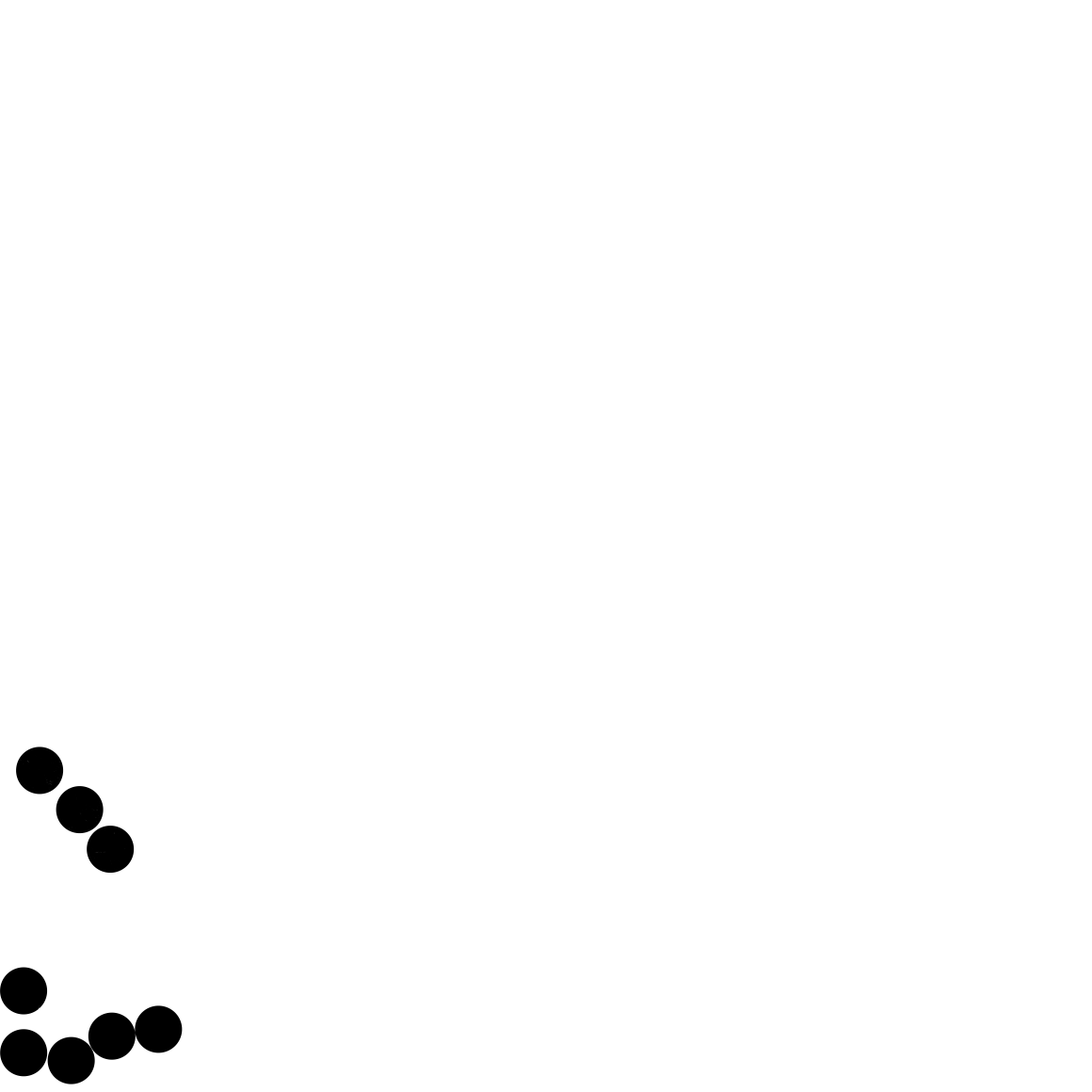}}
  \fbox{\includegraphics[width=1.5in]{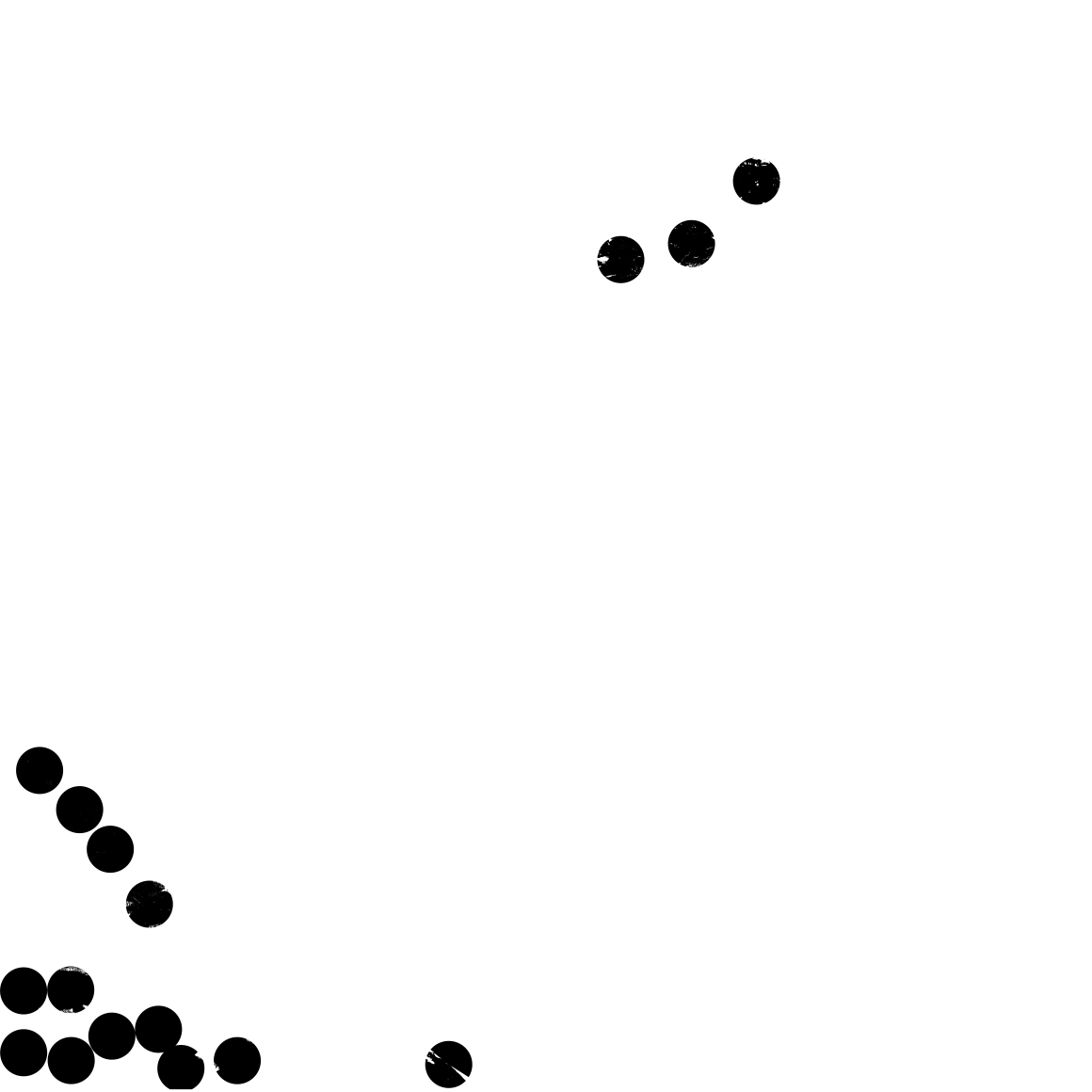}}
  \fbox{\includegraphics[width=1.5in]{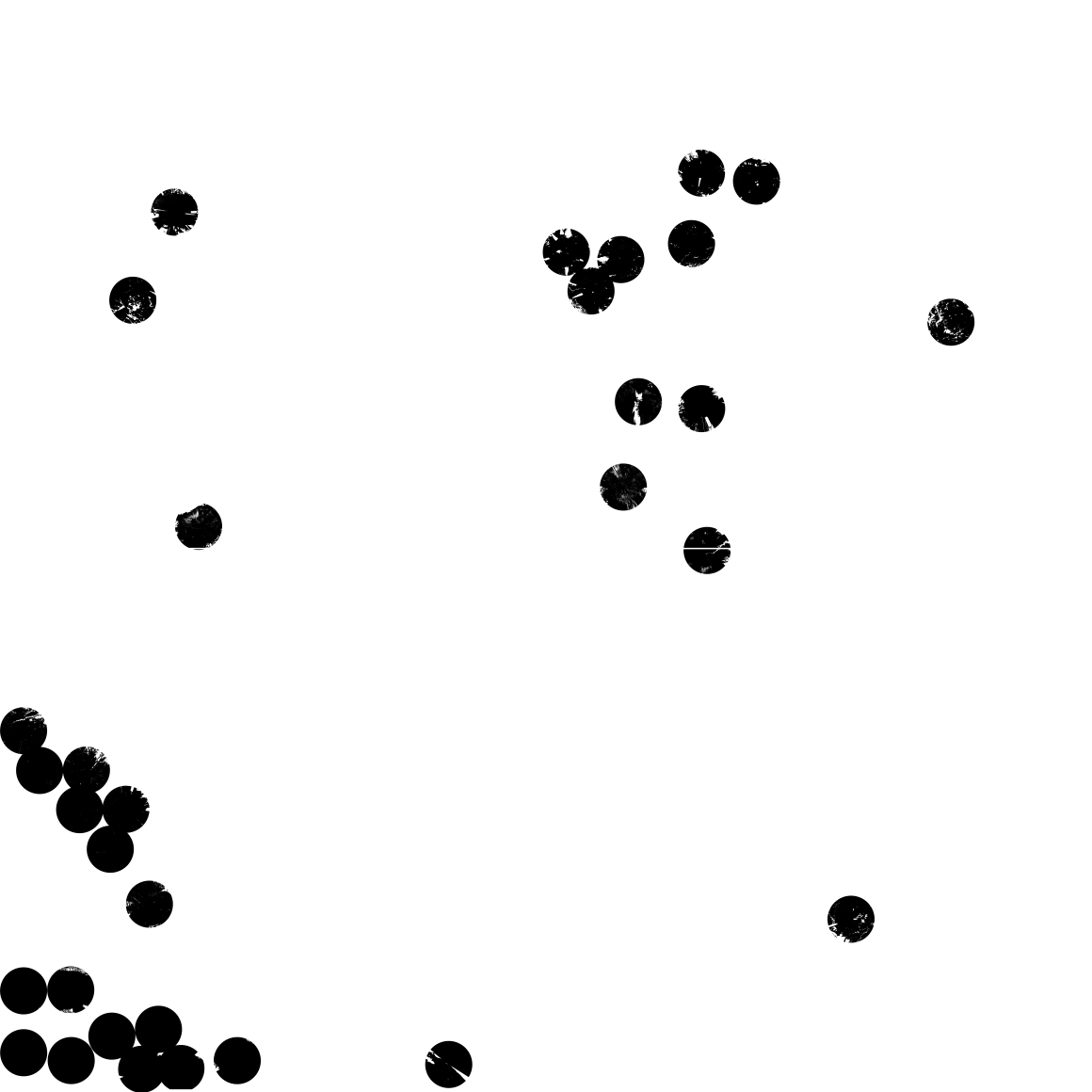}}
  \fbox{\includegraphics[width=1.5in]{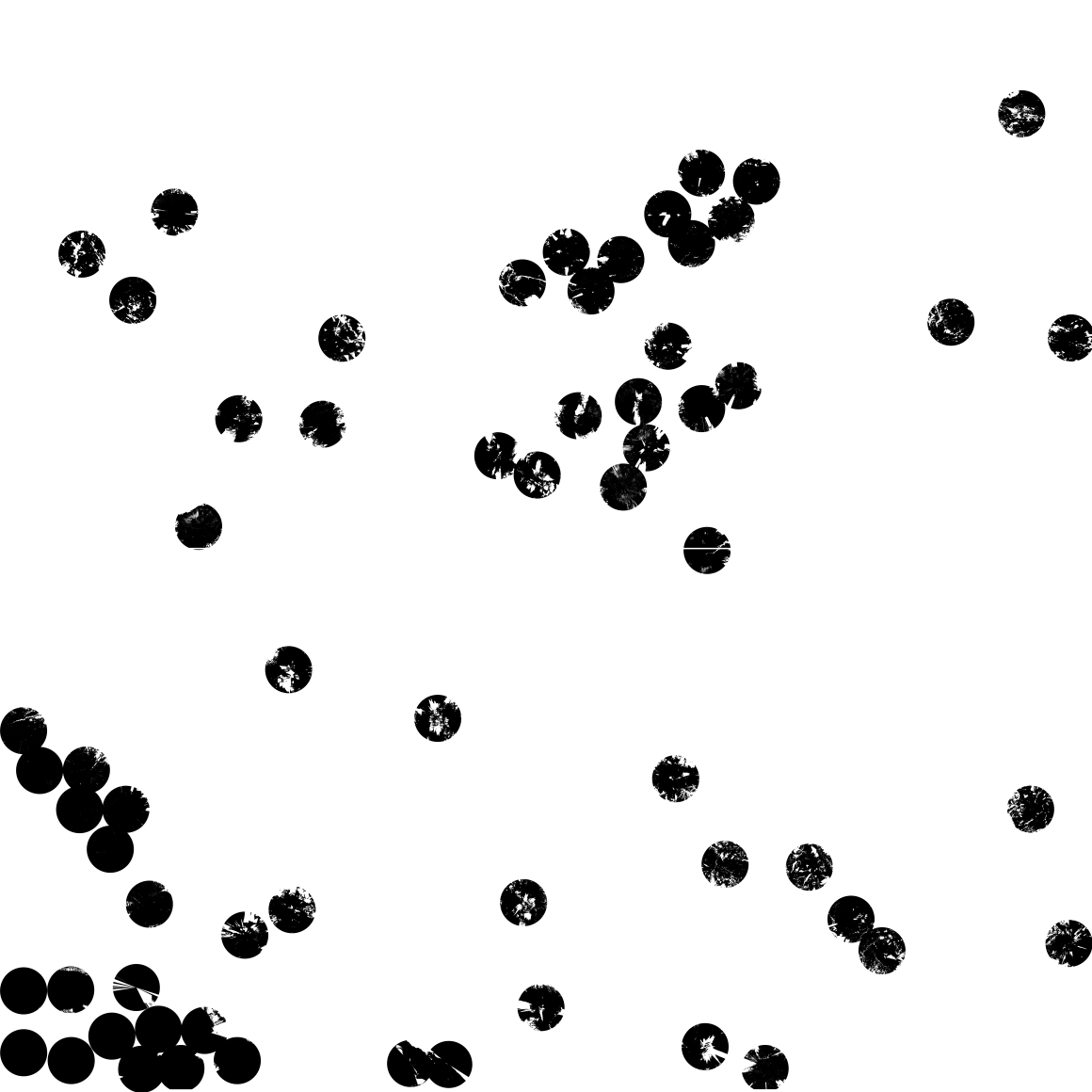}}
  \fbox{\includegraphics[width=1.5in]{figs/usw1-cumshed128.png}}
  \fbox{\includegraphics[width=1.5in]{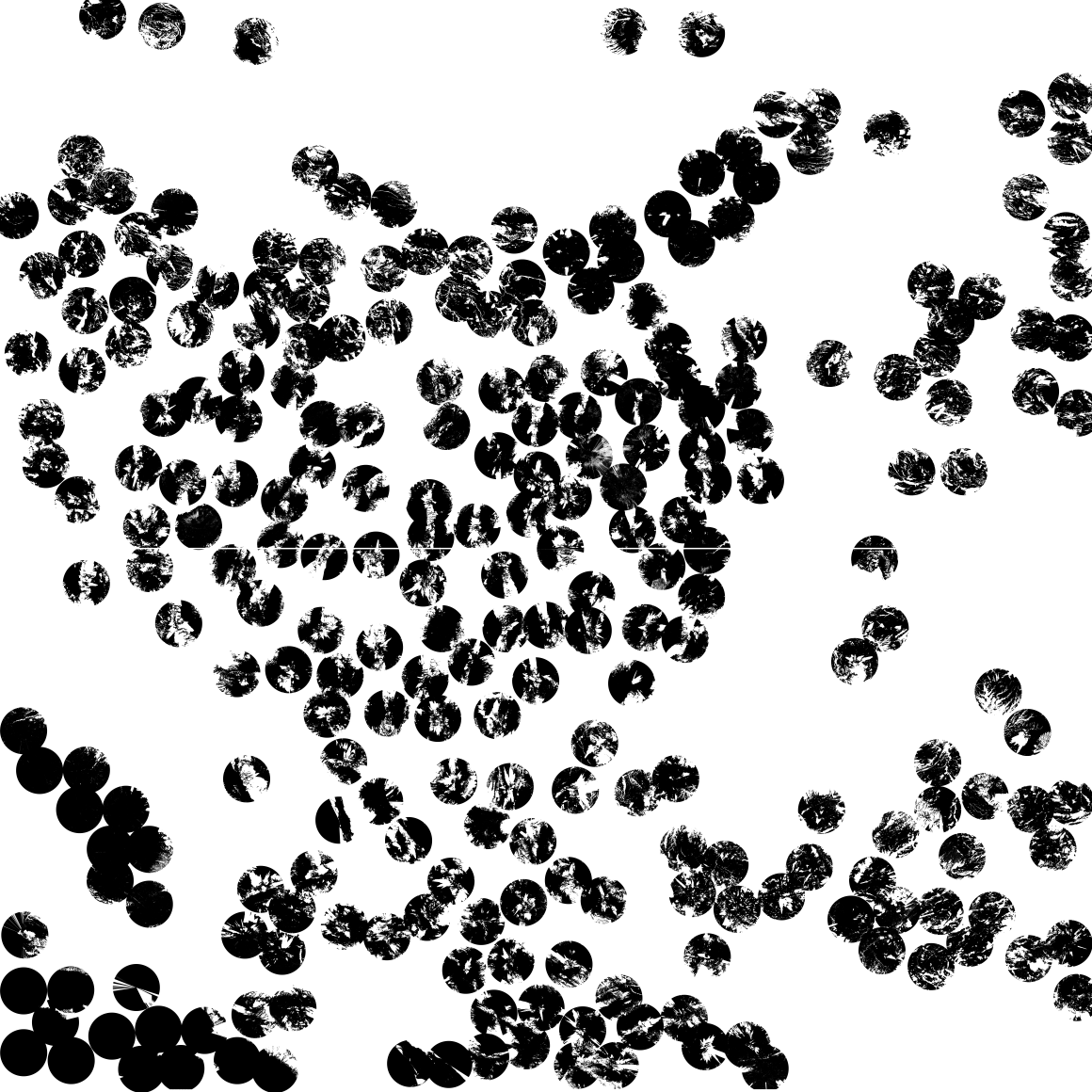}}
  \fbox{\includegraphics[width=1.5in]{figs/usw1-cumshed512.png}}
  \fbox{\includegraphics[width=1.5in]{figs/usw1-cumshed1024.png}}
  \fbox{\includegraphics[width=1.5in]{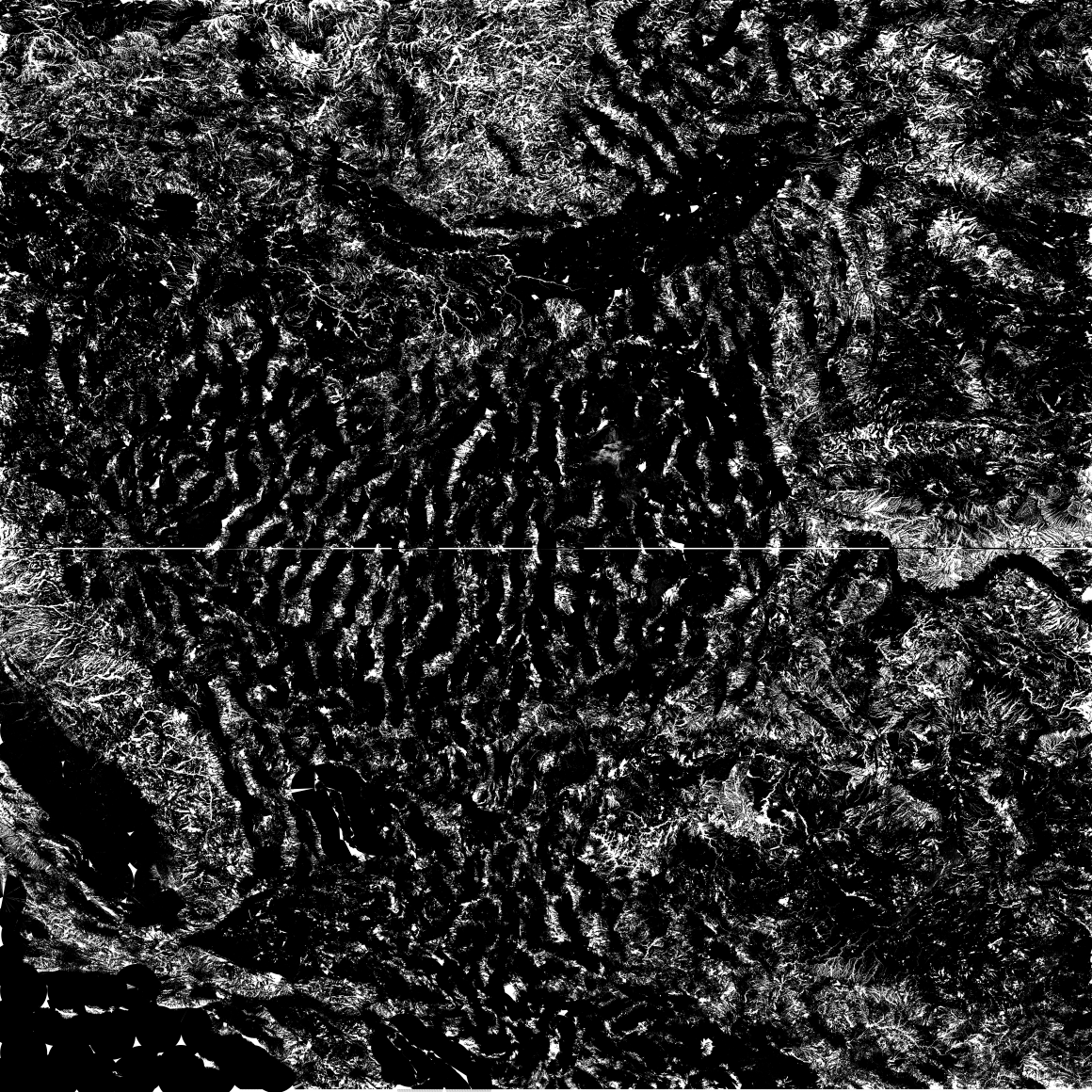}}
  \fbox{\includegraphics[width=1.5in]{figs/usw1-cumshed4096.png}}
  \fbox{\includegraphics[width=1.5in]{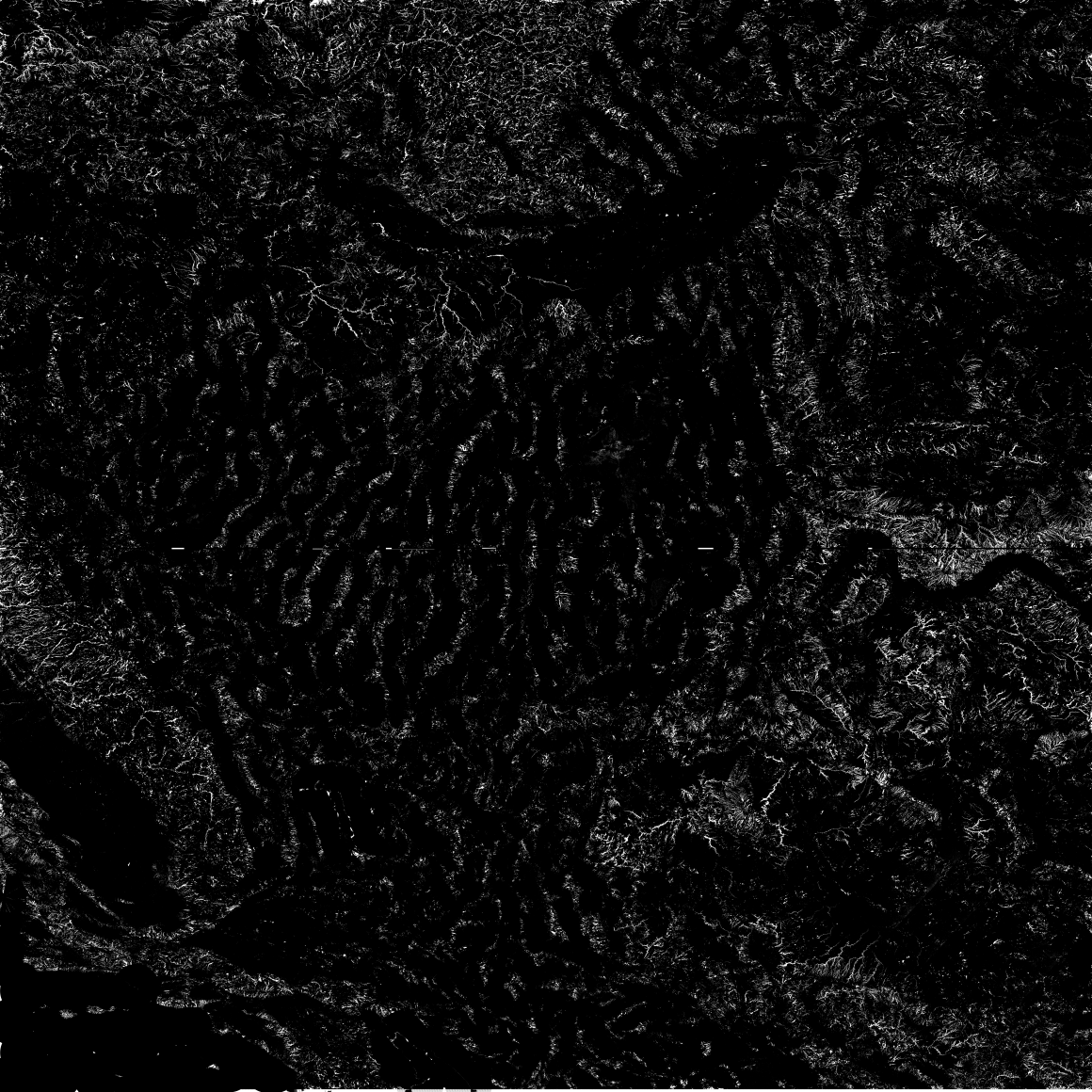}}
  \caption{Cumulative viewsheds for US West after 1, 2, 4, 8, 16, 32, 64, 128, 256, 1024, 2048, 4096, and 5647 transmitters sited}
  \Description{Cumulative viewsheds}
\label{f:usw-sheds}
\end{figure*}

\section{The multiple tranmitter siting process}

This has four stages, summarized below.  For more details, see \citet{wenli-gpu-siting-2016}.

\begin{description}
\item[Vix] finds an approximate visibility index for each possible transmitter location in the terrain, using random sampling.   For each location, i.e., each point in the map, 10 potential receiver locations are chosen uniformly randomly within a circle of radius ROI around the transmitter.  Whether or not each one is visible is computed by testing whether the line of sight between them intersects the terrain.   Extreme accuracy in computing these visible indexes is not required because their only use is to identify potential transmitters.  
  
\item[Findmax] uses those visibility indices to compute a subset of  of the potential transmitters, called \emph{top transmitters}.
  
Merely sorting the potential transmitter list to select the first ones would be wrong.  The problem is there might be a small high visibility region in the terrain.   Inside this region there could be many transmitters, each with a high visibility index, but with largely overlapping viewsheds.   So, they are redundant, but including them in the top list would crowd out lower visibility transmitters that are not redundant and would be useful to include in the solution.

Our solution is to partition the terrain into  blocks of width ROI/3, and select the 20 transmitters in each block.

\item[Viewshed] computes the viewshed of each transmitter in the list returned by Findmax.   It draws a circle of radius ROI around the transmitter and walks around it.   For each point on the circle, it runs a line of sight from the transmitter.   Then it walks along the line of sight, updating a horizon angle, to determine which points interior to the circle are visible.   This process is linear time in the number of points in the circle, i.e., quadratic in the ROI.

  The viewsheds are stored as bitmaps using 64-bit words.

\item[Site] is the heart of the process.   Site greedily determines the set of actual top transmitters.    It maintains a cumulative viewshed bitmap.   At each step, it selects the transmitter, from the set returned by Findmax, whose viewshed would most increase the area of the cumulative viewshed when united with it.   The union process is effected by bitwise operations on the 64-bit words, so it is fast.

  Various optimizations are employed.  E.g., in a later stage, a possible transmitter cannot increase the cumulative viewshed area by more than it would have increased it in an earlier stage.

\end{description}

This paper extends our earlier system to handle much larger datasets---up to two billion elevation posts. 

\section{Implementation}

The above algorithm has been implemented in both serial and parallel versions, using C++ under Linux.   The parallel versions use either OpenMP or CUDA.   The program can run on a server or even on a good laptop, depending on the dataset size.   The total virtual memory used to process one very large terrain was observed to be only 120 bytes per point, although this depends on factors such as the ROI.   The time scales linearly with the relevant parameters, and has a small linear multiplicative factor.   We consider our execution times to be fast enough that we are no longer really concerned with speed, but are testing the maximum feasible terrain size and studying various properties of the process.   This paper's experiments used OpenMP.

We use simple, regular, compact data structures, avoiding recursion, pointers, trees.  This follows the \emph{Structure of Arrays} paradigm.  We avoid the $\log N$ factors in time or space that many other algorithms have; noting that here $N=2^{31}$.  So our total storage is less, execution times small, and processing very large datasets is feasible.

More implementation details are as follows.
OpenMP adds directives to the C++ program so that different iterations of a \emph{for} loop can run in parallel.  This assumes that the different iterations do not affect each other.   E.g., they do not both write to the same variable.   If that is required, then a critical directive can be used to serialize that access.  The resulting program runs on a multicore Intel CPU.   Our usual target machine is a dual 14-core Intel Xeon.   The hard part of programming is designing the algorithm so that the code can be parallelized.


Defining parallel speedup of an algorithm is challenging.  Elapsed real clock time is more useful than CPU time.  A core that is not being used by this algorithm may well not be useful to another simultaneous program because other resources are constrained, such as I/O or memory.    However Xeon CPUs can vary their clock speed over a range of sometime 3:1.   They slow down when idle, but overclock and accelerate when running a compute-bound process.   However, with current integrated circuit technology, the heat generated by a CPU varies with how hard it is computing.   If all the CPU cores are being used, then it might overheat, and so it automatically slows down.   This means that if a  program uses all the cores intensively, they will slow down.  So, even if the program is perfectly parallelizable, the real time speedup will be less than linear.

\section{Testing}

We used 3 test data sets.

\subsection{DEM1000}

This is a trivial test case with only 1,000,000 points; see Figure \ref{f:dem1000-terrain}.   Our laptop runs it in about 5 elapsed seconds, depending on the ROI. Nevertheless, it shows the richness of the cumulative viewsheds; see Figure \ref{f:1000-sheds}.   The stats for that test case are in Table \ref{t:1000}.
  
\subsection{US East}

This dataset has over one billion points.

We generated some terrains using digital elevation models (with a 30-meter resolution) provided by the NASADEM dataset~\cite{NasaDEM}. These data have been recently released by NASA and they were derived from elevations acquired by the Shuttle Radar Topography Mission (SRTM). One of the main advantages of these new models is that cells with missing elevation in the SRTM dataset (i.e., tagged with NODATA) have been filled.

Our US East dataset was extracted from the 1-arc-second NASADEM terrains, and is an example of a relatively flat region.
It has $32,000\times32,000=1\,024\,000\,000$ points.  It bounds are 35N -- 44N (a little less than 44), 85W -- 76W.  Figure \ref{f:usmap} shows the locations of the US West and US East datasets.  Figure \ref{f:useast} shows the US East terrain.  Table \ref{t:useast} summarizes results from some tests on this data.

\subsection{US West}

Our largest test dataset, with over two billion points, is the US-West dataset extracted from the 1-arc-second NASADEM terrains.  It has $46\,400\times46\,400=2\,152\,960\,000$ points.  It bounds are 33N -- 46N (a little less than 46) , 121W -- 108W; see Figure \ref{f:uswest}.  It contains a nice mixture of flat and mountainous terrain.

Figure \ref{f:usw-sheds} shows how the cumulative viewshed progresses as more top transmitters are selected.

\section{Summary and Future Work}

We can process terrains with billions of points to site thousands of radio transmitter towers in $1\frac{1}{2}$ hours, or process terrains with merely a million points in a few seconds.   Future work is to get the GPU code working on these large example, and experiment on the sensitivity of the result to lowered accuracy in the data.

\bibliographystyle{ACM-Reference-Format}
\bibliography{wrf,bibliography}

\end{document}